\documentclass[sigconf]{acmart}
\usepackage{bbm}
\usepackage{multirow}
\usepackage{colortbl}
\usepackage{xcolor}
\usepackage{array}
\usepackage{subfigure}

\AtBeginDocument{%
  }

\setcopyright{acmlicensed}
\copyrightyear{2018}
\acmYear{2018}
\acmDOI{XXXXXXX.XXXXXXX}

\usepackage[ruled,linesnumbered]{algorithm2e}
\newtheorem{theorem}{Theorem}
\newtheorem{Def}{Definition}

\begin{document}

\title{Towards Fair Graph Representation Learning in Social Networks}



\author{Guixian Zhang}
\orcid{0000-0002-7632-8411}
\author{Guan Yuan}
\orcid{0000-0003-3148-9817}
\affiliation{%
	\institution{School of Computer Science and Technology, Mine Digitisation Engineering Research Center of the Ministry of Education, China University of Mining and Technology}
	\city{Xuzhou}
	\state{Jiangsu}
	\country{China}
}

\author{Debo Cheng}
\orcid{0000-0002-0383-1462}
\author{Lin Liu}
\orcid{0000-0003-2843-5738}
\author{Jiuyong Li}
\orcid{0000-0002-9023-1878}
\affiliation{\institution{UniSA STEM, University of South Australia}
	\city{Adelaide}
        \state{SA}
	\country{Australia}}

\author{Shichao Zhang}
\orcid{0000-0003-4052-1823}
\affiliation{%
	\institution{Guangxi Key Lab of Multisource Information Mining \& Security,
Guangxi Normal University}
	\city{Guilin}
	\state{Guangxi}
	\country{China}
}



\renewcommand{\shortauthors}{Guixian et al.}

\begin{abstract}
    With the widespread use of Graph Neural Networks (GNNs) for representation learning from network data, the fairness of GNN models has raised great attention lately. Fair GNNs aim to ensure that node representations can be accurately classified, but not easily associated with a specific group.  Existing advanced approaches essentially enhance the generalisation of  node representation in combination with data augmentation strategy, and do not directly impose constraints on the fairness of GNNs. In this work, we identify that a fundamental reason for the unfairness of GNNs in social network learning is the phenomenon of \textit{social homophily}, i.e., users in the same group are more inclined to congregate. The message-passing mechanism of GNNs can cause users in the same group to have similar representations due to social homophily, leading model predictions to establish spurious correlations with sensitive attributes. Inspired by this reason, we propose a method called \textbf{E}quity-\textbf{A}ware \textbf{GNN} (\textbf{EAGNN}) towards fair graph representation learning. Specifically, to ensure that model predictions are independent of sensitive attributes while maintaining prediction performance, we introduce constraints for fair representation learning based on three principles: sufficiency, independence, and separation. We theoretically demonstrate that our EAGNN method can effectively achieve group fairness. Extensive experiments on three datasets with varying levels of social homophily illustrate that our EAGNN method achieves the state-of-the-art performance across two fairness metrics and offers competitive effectiveness.
\end{abstract}



\maketitle

\section{Introduction}
Graph Neural Networks (GNNs) have emerged as a powerful class of machine learning models, particularly suited for capturing complex relationships and interactions in a wide range of real-world systems~\cite{wu2024graph, wang2022unbiased, huang2024can}. GNNs update node representations by aggregating and transforming information from neighbouring nodes, a process commonly referred to as message-passing~\cite{xu2018how}. The message-passing mechanism allows the model to capture both the characteristics of individual nodes and their connectivity patterns within the graph, thus giving GNNs a powerful performance on various downstream tasks. Despite their strong performance, GNNs are often criticized for issues related to fairness and trustworthiness. Specifically, GNNs may inadvertently learn and amplify biases in the training data~\cite{mehrabi2021survey}, meaning that any inherent biases in the data will be reflected in the model’s predictions, potentially leading to unfair decisions for certain groups. These biased predictions raise significant ethical and social concerns, particularly in real-world applications like recommender systems~\cite{xuan2023knowledge}, rumour detection~\cite{zhang2023rumor}, and social bot detection~\cite{zhang2024bayesian}, where fairness is critical.

Fairness challenges in GNNs differ from those in other machine learning models because graph data involves not only node features but also the structure of the graph itself~\cite{xu2018how}. However, in social networks, users’ interactions are influenced by sensitive attributes, such as gender, age and race, which can introduce biases into the graph structure. This phenomenon, referred to as ``social homophily'', describes the tendency of individuals to form connections with others who are similar to them~\cite{ma2023group, jiang2024challenging, stoica2024fairness}. This scenario is also summarised by the phrase ``similarity breeds connection''~\cite{mcpherson2001birds}. For example, Stoica et al.~\cite{stoica2018algorithmic} found that social media users are more likely to connect with others in the same age group, with male users displaying stronger homophily than female users. In social recommender systems, if users in a same group are frequently observed connecting with each other, the model may record and amplify this behaviour, ultimately recommending friends only within the same group, thereby causing bias~\cite{wang2022unbiased}.  To the best of our knowledge, no prior work has explicitly  defined the problem of social homophily in fair graph representation learning, nor provided a practical solution for addressing its impact on fairness.


\begin{figure}[t]
    \centering
    \includegraphics[width=0.56\linewidth]{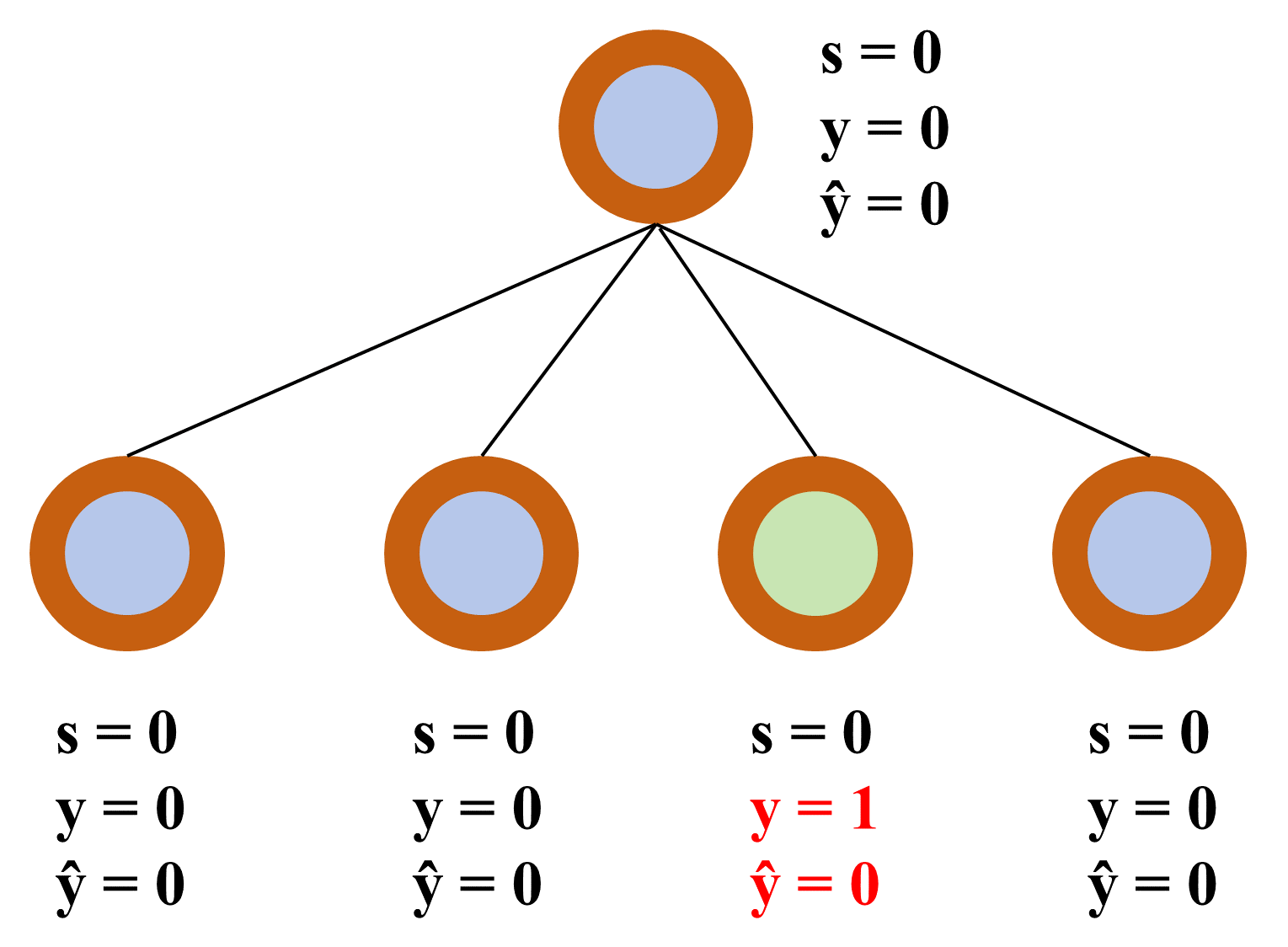}
    \caption{An example of graph data is provided to illustrate the social homophily. $S$ denotes the sensitive attribute, $Y$ denotes the label, and $\hat{Y}$ denotes the prediction. }
    \label{motivation}
\end{figure}

 Expectations of fairness can be divided into three criteria~\cite{mehrabi2021survey, sohn2024fair}, and social homophily affects the fairness of GNNs from these three perspectives as well: sufficiency, independence, and separation.  To show social homophily in social network, we present an example in Figure~\ref{motivation}. In this social network, five nodes belonging to the same group are clustered together, demonstrating high social homophily. In such a case, the neighbourhood  aggregation mechanism of a GNN can easily treat group membership as a key factor when predicting node labels. For instance, all nodes with $s = 0$ (sensitive attribute) may be predicted to have $\hat{y} = 0$, resulting in fairness issue.  Firstly, we infer that the different nodes are not sufficiently learned, making the model overly concerned with sensitive attributes (i.e., violating sufficiency) during the training. In this case, the model may build shortcuts for sensitive attributes and node labels in training, leading to spurious correlations (i.e., violating independence). Even when model evaluations are controlled within a particular category, the model exhibits different error rates in its predictions based on the sensitive attributes of the individuals (i.e., violating separation). While existing work has partially explored the impact of social homophily~\cite{dong2022edits,wang2022improving,zhang2023fpgnn, guo2023towards, he2023fairmile,li2024rethinking,zhang2024learning}, it often fails to clearly define the problem. Moreover, these methods focus on improving fairness through techniques like graph rewiring or graph generation, which primarily enhance the generalisation of node representations rather than directly addressing fairness concerns.

To address the above-mentioned challenges, in this work, we formally define social homophily in graph data and propose a method, referred to as \textbf{E}quity-\textbf{A}ware \textbf{GNN} (\textbf{EAGNN}) to overcome the effect of social homophily.  Specifically,  to reduce the influence of sensitive attributes on model predictions, we design loss functions of our EAGNN based on the the requirements of \textit{sufficiency}, \textit{independence} and \textit{separation} to serve as fairness constraints. We theoretically demonstrate that social homophily leads to a model that clearly identifies groups of nodes, and that sufficiency requires node representations are sufficiently trained across populations thus avoiding this issue. Independence requires that sensitive attribute $S$ be independent of the prediction $C(\mathbf{H})$, i.e., $S \perp C(\mathbf{H})$, where $C$ is the classifier, and $\mathbf{H}$ is the node representation. Separation involves conditional independence, defined as $S \perp C(\mathbf{H}) \mid Y$, where $Y$ is the label. We theoretically demonstrate that the designed loss function can achieve group fairness regarding independence and separation.  These constraints complement each other and help achieve a balance between accuracy and fairness. In summary, our main contributions are as follows:
\begin{itemize}
    \item   We identify that bias in GNNs can be explained by social homophily and demonstrate its effects theoretically, providing a new perspective for analysing fairness in graph learning. 
    \item  We propose a novel method, EAGNN, which overcomes the effects of social homophily through loss functions in three important perspectives: sufficiency, independence, and separation. We theoretically demonstrate that our EAGNN achieve group fairness. 
    \item  We conducted extensive experiments on three datasets with varying degrees of social homophily, and the results show that EAGNN achieves excellent performance in terms of both effectiveness and fairness.
\end{itemize}

\section{Preliminary}
In this section, we collate the notations used in this paper and then define social homophily and fairness metrics.

\subsection{Notations}
In our study, we employ the notation $\mathcal{G}=(\mathcal{V}, \mathcal{E}, \mathbf{A}, \mathbf{X})$ to denote an attribute graph. Here, $\mathcal{V}= \left\{v_{1}, \ldots, v_{n}\right\}$ represents the node set, $\mathcal{E}$ denotes the edge set, $\mathbf{A} \in \mathbb{R}^{n \times n}$ signifies the adjacency matrix, and $\mathbf{X} \in \mathbb{R}^{n \times d}$ corresponds to the matrix of node attributes. The quantity $n$ denotes the number of nodes, while $d$ indicates the attribute dimension. Each node $v_{i}$ is characterized by a label $y$ and a sensitive attribute $s_i$, with $s_i$ and $y_i$ both belonging to the set $\{0, 1\}$. The objective of fair representation learning is to develop a model capable of delivering accurate predictions without being influenced by sensitive attributes. In the context of binary classification, utilizing a classifier $C$ and a representation $\mathbf{h}_{i}$ for each node $v_{i}$, we derive the predicted outcome $\hat{y}_i = C(\mathbf{h}_{i})$. The performance and fairness of the model are assessed by quantifying the relationship between $\hat{y}_i$, the actual target labels $y_i$, and the sensitive attribute $s_i$.

\subsection{Causal View}

\begin{figure}[t]
    \centering
    \includegraphics[width=0.365\linewidth]{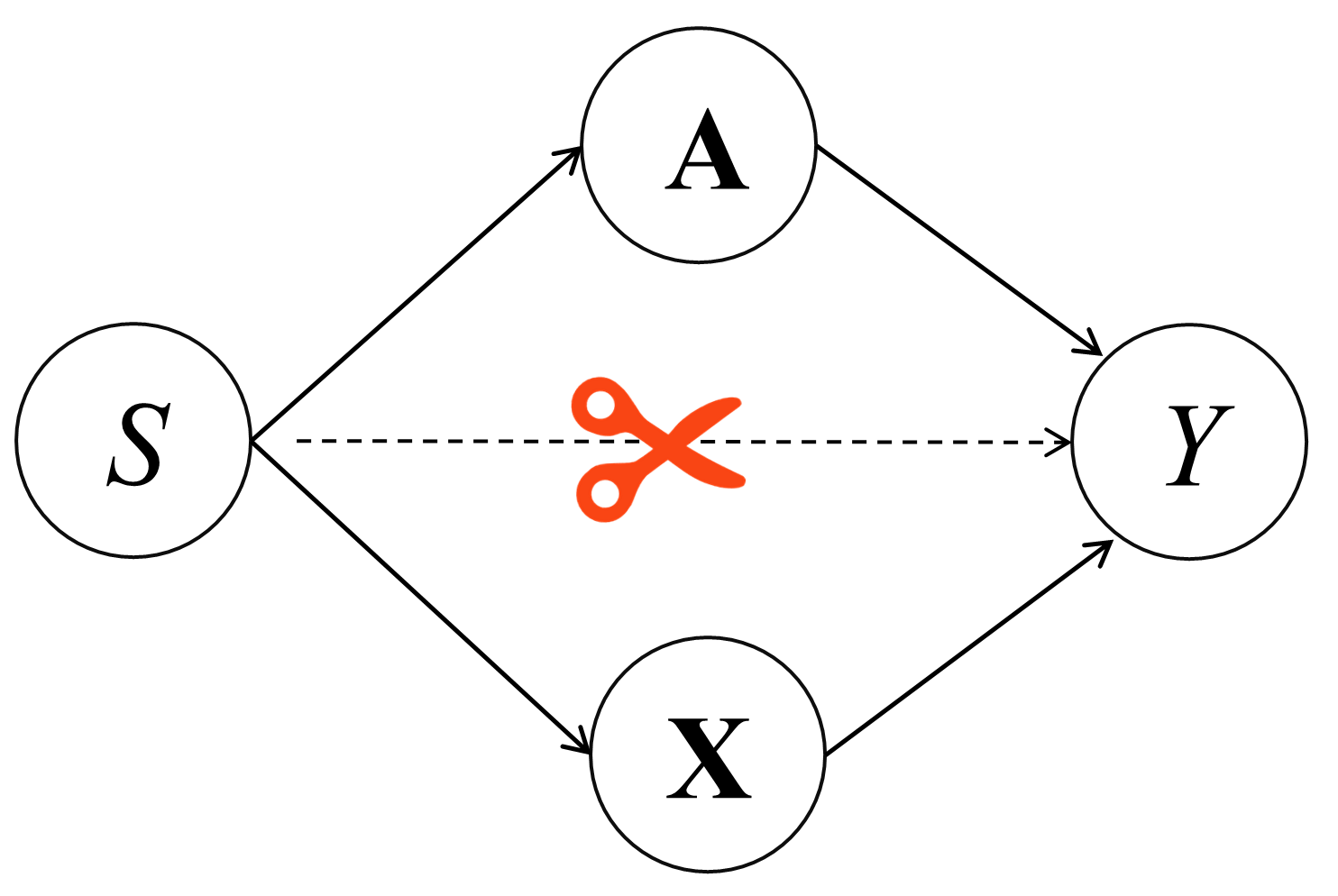}
    \caption{The proposed SCM for representing the graph data generation process. We aim to avoid building spurious correlations between $S$ and $Y$ during the training process.}
    \label{causal}
\end{figure}
In this paper, we propose use the Structure Causal Model (SCM)~\cite{pearl2009causality} as shown in Figure~\ref{causal} to represent the underlying process of the graph data generation and illustrate the motivation behind our work. During the generation of graph data, the sensitive (bias) variable $S$ is typically unobserved, and affects the observed node attributes $\mathbf{X}$ and topology $\mathbf{A}$. For instance, people of different races had varying positivity rates during COVID-19 because healthcare access may have been influenced by their economic status~\cite{martinez2020sars, chen2022impact}. In such cases, training a GNN can easily establish spurious correlations $S$ and $Y$. Therefore, in our work, we design constraints based on the principles of sufficiency, independence, and separation to prevent $S$ from affecting the model's prediction $\hat{Y}$, while fully considering social social homophily.

\subsection{Social Homophily}

In this work, we focus on improving the fairness of GNNs by reducing the effect of social homophily. Similar to the definition of homophily~\cite{zhu2020beyond, ma2022homophily}, we define social homophily in the graph based on whether connected node pairs belong to the same group: 
\begin{Def}[Social Homophily]
\label{socialhomo}
Let $\mathcal{G}$ be a graph and $S$ the set of node group labels (sensitive attribute). The social homophily ratio, denoted as $SH(\mathcal{G}, S)$, is the proportion of edges connecting nodes within the same group. It is given by:
\begin{equation}
    SH(\mathcal{G}, S) = \frac{1}{|\mathcal{E}|} \sum_{(j, k) \in \mathcal{E}} \mathbbm{1}(s_j = s_k),
\end{equation}
where $|\mathcal{E}|$ is the total number of edges and $\mathbbm{1}$ denotes the indicator function.
\end{Def}

A graph is considered to have a high degree of social homophily when $SH(\cdot)$ is large (typically, $0.5 \le  SH(\cdot) \le  1$). Indeed, if a graph belongs to a social network, then its social homophily should be greater than 0.5. Note that in our experiments, to better validate the effectiveness of EAGNN, we performed experimental validation on two datasets with high social homophily (greater than 0.8) and one dataset with relatively low social homophily (less than 0.8).

\subsection{Fairness Metric}
In this paper, we use two group-specific fairness metrics to evaluate the fairness of GNNs.

\begin{Def}[Statistical Parity~\citep{dwork2012fairness}] Statistical parity stipulates that the proportion of individuals receiving positive classifications should be approximately equal across demographic groups, i.e., $S \perp \hat{Y}$.
\end{Def}

\begin{Def}[Equal Opportunity~\citep{hardt2016equality}] Equal opportunity stipulates that the true positive rate should be approximately equal across demographic groups, i.e. $S \perp \hat{Y} \mid Y$. 
\end{Def}


In this paper, we use $\Delta_{mathrm{SP}}$ and $\Delta_{mathrm{EO}}$ to measure the statistical parity difference and the equal opportunity difference between two groups, respectively, i.e., the smaller the value is, the closer the group fairness is. Specifically, for a specific node $v$, which has a sensitive attribute $s$, a predicted outcome $\hat{y}$, and a label $y$, then according to Definition 1 and Definition 2, we can compute the fairness metrics for this node:
\begin{equation}
	\Delta_{\mathrm{SP}}=|P(\hat{y}=1 \mid s=0)-P(\hat{y}=1 \mid s=1)|, 
\end{equation}
\begin{equation}
	\Delta_{EO} =|P(\hat{y}=1 \mid y=1, s=0)- P(\hat{y}=1 \mid y=1, s=1)|.
\end{equation}

\section{The Proposed EAGNN Method}
In this section, we start by by describing how a GNN model makes predictions, and the fairness issue from three perspective: sufficiency, independence and separation.  We then design constraints for fair predictions from the three perspectives. 

\subsection{Encoding and Classification}
GNNs operate on graph data by propagating information between neighbour nodes. In a GNN, the representation vector $\mathbf{h}_{i}^{k}$ of node $v_{i} \in \mathcal{V}$ at the $k$-th layer captures the structural information within the $k$-hop subgraph surrounding $v_{i}$. The update process for the $k$-th layer of a GNN is formally defined as:
\begin{equation}
\mathbf{h}_{i}^{(k)}=\operatorname{Update}\left(\mathbf{h}_{i}^{(k-1)} \text {, } \operatorname{Aggregate}\left(\left\{ \mathbf{h}_{u}^{(k-1)} \mid u \in \mathcal{N}(v_i) \right\}\right)\right),
\end{equation}
where ${N}(v_i)$ is the set of neighbours of $v_i$. 

After obtaining the node representation $\mathbf{H}$,   a MultiLayer Perceptron (MLP) is used to serve as a classifier $C$ for predicting $\hat{Y}$:
\begin{equation}
    \hat{Y} = C(\mathbf{H}).
\end{equation}

Specifically, the training loss function of the classifier is expressed as follows:
\begin{align}
\mathcal{L}_{C}  = - \mathbb{E}_{v_{i} \sim \mathcal{V}} \left(y_{i} \log \left({\hat{y}_{i}}\right) -  \left(1-y_{i}\right) \log \left(1-{\hat{y}_{i}}\right)\right),
\end{align}
where $y_{i}$ is the label of the node $v_i$.

\subsection{Sufficiency}
We assume the features of node $v_i$ are sampled from the feature distribution $\mathcal{F}_{s_{i}} $, i.e., $\mathbf{x}_i \sim \mathcal{F}_{s_{i}}$, with $ \mu(\mathcal{F}_{s_{i}})$ denoting the mean of $\mathcal{F}_{s_{i}}$. The features are independent and the magnitude of each feature in $\mathbf{X}$ does not exceed a predefined scalar bound $B$, i.e., $\max_{i, j} |\mathbf{X}[i, j]| \leq B$.  Based on these assumptions, we 
derive Theorem~\ref{th1}. The proof of the Theorem~\ref{th1} is provided in Appendix~\ref{proof1}.


\begin{theorem}
\label{th1}
Let $\mathcal{G}$ be a graph defined by ${\mathcal{V}, \mathcal{E}}$. Each node  $v_i$  in  $\mathcal{G}$ is characterised by a feature vector  $\mathbf{x}_i \in \mathbb{R}^l$ and a sensitive attribute $s_i$. For any node $ v_i \in \mathcal{V}$  belonging to group $b$, the expectation of the pre-activation output of a single Graph Convolutional Network (GCN) operation is given by:
\begin{equation}
    \mathbb{E} \left[ \mathbf{h_i} \right] = \mathbf{W} \left( \mathbb{E}_{b \sim \mathcal{D}_{s_{i}}, \mathbf{x} \sim \mathcal{F}_{b}} \left[ x \right] \right), 
\end{equation}
where $\mathbf{W}$ is the parameter matrix in the GCN and $ \mathcal{D}_{s_{i}}$ is the neighbour distribution.

Moreover, for any positive scalar $t$, the likelihood that the Euclidean distance between the actual output $\mathbf{h}_i$, and this expected output exceeds $t$ is upper-bounded by:
\begin{equation}
     \mathbb{P} \left( \left\| \mathbf{h_i} - \mathbb{E} \left[ \mathbf{h_i} \right] \right\|_{2} \geq t \right) \leq 2 \cdot l \cdot \exp \left( -\frac{\operatorname{deg}(v_i) t^{2}}{2 \rho^{2}(\mathbf{W}) B^{2} l} \right)
\end{equation}
where $l$ denotes the feature dimensionality, and $\rho(\mathbf{W})$ denotes the largest singular value of $\mathbf{W}$.
\end{theorem}

By Theorem~\ref{th1}, we observe that the GNN model will map nodes with the same sensitive attribute to an expectation-centred area in the embedding space, with a small distance. This implies that the node representations in this case have a strong correlation with the sensitive attributes, which inevitably makes the node representations pay too much attention to the sensitive attributes during the training process. To achieve sufficiency, we need to enhance the learning of node representations that belong to different sensitive groups but share similar attributes. Specifically, we first select the nodes that have not been sufficiently learned:
\begin{equation}
    M_{i}=\left\{\begin{array}{ll}
1, & \text { if } \operatorname{sim}\left(x_{i}, x_{j}\right)>\theta \text { and }\left(s_{i} \neq s_{j}\right) \\
0, & \text { otherwise }
\end{array}\right.
\end{equation}
where $v_j$  is any node in the graph, $\operatorname{sim}\left(x_{i}, x_{j}\right)$ is the similarity between nodes  $v_i$  and  $v_j$, $\theta$  is a predefined threshold, $s_{i}$  and  $s_{j}$  are the sensitive values of nodes  $i$  and  $j$, respectively.

Next, the sufficiency loss  $L_{suff}$  can be expressed as:
\begin{equation}
\mathcal{L}_{suff}= - \mathbb{E}_{v_{i} \sim \mathcal{V}} \left(\frac{1}{2}\left(\hat{y_i} - y_i\right)^{2} \cdot M_{i}\right).    
\end{equation}

With $\mathcal{L}_{suff}$, we can make the model focus on nodes that belong to different groups but have high attribute similarity during the training, thus avoiding overfitting the model to sensitive attributes.

\subsection{Independence}
The independence condition requires that $S$ be independent of $C(\mathbf{H})$, i.e., $S \perp C(\mathbf{H})$. In this section, we choose statistical parity as the criterion for independence and randomly generate $S'$ to quantify the fairness level using the discriminator $D$ which is constructed by a MLP. This design ensures that the predictive output of the model remains consistent under different values of the sensitive attribute, thereby reducing prediction bias with respect to the the sensitive attribute. Specifically, to achieve independence, we introduce an independence penalty $\mathcal{L}_{in}$ for classifier $C$:
\begin{equation}
\mathcal{L}_{in} = \mathbb{E}_{v_{i} \sim \mathcal{V}}(\log D(\hat{y}_{i}, s_i) + log (1 - D (\hat{y}_{i}, s'_i)),
\end{equation}
where $D_{\epsilon }$ is modeled by a MLP and $s'_i \in S'$ is the randomly generated sensitivity value.

Through Theorem~\ref{th2}, we can obtain the optimal $D^{*}$. The proof of the Theorem~\ref{th2} is offered in Appendix~\ref{proof2}.

\begin{theorem}
\label{th2}
Let $p_{S}$ and $p_{\hat{Y}}$ represents the marginal density functions of the random variable $S$ and $\hat{Y}$, respectively. $p_{\hat{Y}\mid S}$ is the conditional density function of $\hat{Y}$ given $S$, and $p_{\hat{Y}, S}$ is the joint density function of $\hat{Y}$ and $S$. Now, we introduce the discriminator $D$ to determine whether the model outputs $\hat{Y} = C(\mathbf{H})$ are independent of $S$. The optimal discriminator $D^{*}$, which maximises the objective function $\mathcal{L}_{in}$ over all possible discriminators $D$, can be expressed as:
\begin{equation}
p_{\hat{Y}, S}(\hat{y}, s)=p_{\hat{Y}}(\hat{y}) p_{S}(s),
\end{equation}
where $\hat{y} \in \hat{Y}$ and $s \in S$.
\end{theorem}

The Independence constraint requires that the marginal distribution of the model output $\hat{Y}$ does not change given the sensitive attribute $S$. Theorem~\ref{th2} theoretically verifies that $\mathcal{L}_{in}$ captures the difference between $P(\hat{Y}\mid S)$ and $P(\hat{Y})$, providing a theoretical justification for using $\mathcal{L}_{in}$ to achieve statistical parity, i.e., $P(\hat{y} = 1 \mid s = 1) = P(\hat{y} = 1 \mid s = 0)$. We can optimise $C$ in a fairness-conscious way by incorporating the additional penalty $\mathcal{L}_{in}$ into the fair risk minimisation problem. The proof of the Theorem~\ref{th2} is provided in Appendix~\ref{proof2}.

\subsection{Separation}
The goal of separation is to ensure $S \perp C(\mathbf{H}) \mid Y$, i.e., $S \perp \hat{Y} \mid Y$, but in real-world applications, the joint distribution of the sensitive attribute $S$ and the label $Y$ may not be uniform, leading to certain combinations $S$ and $Y$ occurring more frequently in the data. To address this, we introduce an  $\epsilon$ function as a density ratio estimator in the separation constraint. This function adjusts for the non-uniformity of the distribution, ensuring that each combination is fairly considered when estimating the conditional probabilities. Specifically, the separation constraint is designed as follows: 
\begin{equation}
R_{se} = \mathbb{E}_{v_{i} \sim \mathcal{V}}(\log D(\hat{y}_{i}, s_i, y_i) + \epsilon(s',y)log (1 - D (\hat{y}_{i}, s'_i, y_i)),
\end{equation}
where $s'_i \in S'$ is the randomly generated sensitivity value.

\begin{theorem}
\label{th3}
Let $p_{\hat{Y} \mid Y}$ be the conditional density function of $\hat{Y}$ given $Y$ and $p_{\hat{Y} \mid S,Y}$ be of given $Y$ and $S$. We are interested in finding the optimal discriminator $D^{*}$ that maximizes a certain objective function $R_{se}$ over all possible discriminators $D$: 
\begin{equation}
\frac{p_{\hat{Y} \mid S, Y}(\hat{y} \mid s, y)}{p_{\hat{Y} \mid S, Y}(\hat{y} \mid s, y)+\epsilon(s, y) p_{\hat{Y} \mid Y}(\hat{y} \mid y) \frac{p_{S^{\prime}, Y}(\hat{y}, y)}{p_{S, Y}(\hat{y}, y)}},  
\end{equation}
for all $\hat{y} \in \hat{Y}$, $s \in S$, and $y \in Y$, where $p_{S',Y}$ and $p_{S,Y}$ are the joint density functions of $S'$ and $Y$ and of $S$ and $Y$, respectively.
\end{theorem}

If $\epsilon(s, y) p_{S^{\prime}, Y}(s, y) = p_{S, Y}(s, y)$, then from Theorem 1 of ~\cite{goodfellow2014generative}, we can infer that $R_{se}$ can be explained by the Jensen-Shannon divergence $JSD(\cdot,\cdot)$. Thus, we have: 
\begin{equation}
R_{se} = 2*JSD\left(P(\hat{Y}, Y, S), P(\hat{Y},Y) P(S)\right) -  \log 4,
\end{equation}

If $JSD(\cdot)= 0$, it implies $P(\hat{Y}\mid Y, S)= P(\hat{Y}\mid Y)$.In other words, the predicted probability is the same for a specific group, regardless of the sensitive attribute, i.e., $P(\hat{y} \mid y=1, s=0)= P(\hat{y}=1 \mid y=1, s=1)$. 

To achieve $\epsilon(s, y) p_{S^{\prime}, Y}(s, y) = p_{S, Y}(s, y)$, we need to fit a density-ratio estimator $\epsilon$ by maximising:
\begin{equation}
R_{\epsilon} = E_{S,Y}[\log D_{\epsilon }(S, Y)] + E_{S,Y}[\log(1 - D_{\epsilon }(S', Y))],    
\end{equation}
where $D_{\epsilon }$ is modeled by a MLP. 

The optimal decision function $D_{\epsilon }^*$ is defined as:
\begin{equation}
    D_{\epsilon }^* = \mathop{\arg\max}\limits_{\epsilon} R_{\epsilon}(D_{\epsilon }),
\end{equation}
where $R_{\epsilon}(D_{\epsilon })$ represents the reward associated with the decision function $D_{\epsilon }$ parameteris ed by $\epsilon$. Under this optimal decision function, it holds that:
\begin{equation}
\frac{D_{\epsilon }(s, y)}{1 - D_{\epsilon }(s, y)} = \frac{p_{S,Y}(s, y)}{p_{S',Y}(s, y)},    
\end{equation}
where $D_{\epsilon }(s, y)$ is the decision function output for a given $s$ and $y$, and \(p_{S,Y}(s, y)\) and \(p_{S', Y}(s, y)\) represent the joint probability distributions of the sensitive attribute and outcome under different scenarios \(S\) and \(S'\), respectively.  This ensures that the decision-making process is balanced in terms of opportunities across different scenarios.

By applying Theorem~\ref{th3}, we ensure the implementation of equal opportunity in our proposed EAGNN method. The proof of the Theorem~\ref{th3} is provided in Appendix~\ref{proof3}. Thus, based on $R_{\beta}$ and $R_{se}$,  the final separation constraint is given by:

\begin{equation}
 \mathcal{L}_{se} = R_{\epsilon} + R_{se}.
\end{equation}

\subsection{Model Training}

The proposed EAGNN  method improves model fairness by integrating multiple fairness constraints while mitigating the social homophily present in the graph. The core of our EAGNN is to combine the classification loss ($\mathcal{L}_{C}$) with three key fairness constraints during model training, resulting in a weighted composite loss function:

\begin{equation}
    \mathcal{L} = \mathcal{L}_{C} + \alpha * \mathcal{L}_{suff} + \beta * \mathcal{L}_{in}  +  \gamma * \mathcal{L}_{se}, 
\end{equation}
where $\alpha$, $\beta$ and $\gamma$ are the weights assigned to each fairness constraint. Specifically, we balance the model's predictive performance and fairness by adjusting the three weights of each loss term, aiming to reduce the model's bias toward specific groups without sacrificing too much effectiveness.

\section{Experiments}
In this section, we conduct extensive experiments to evaluate the effectiveness of the EAGNN method and to assess the importance of each component.

\subsection{Experiment setup}

\subsubsection{Real-world datasets}
\begin{table}[htp]
\centering
	\caption{A summary of the datasets.}
	\begin{center}
		\begin{tabular}{llll}
			\hline
			Dataset           & Credit  & German   & Bail \\
			\hline
			\#of nodes         & 30,000  & 1,000                & 18,876        \\
			\#of node attributes  & 13    & 27                  & 18              \\
			\#of edges        & 1,436,858   & 22,242             & 321,308      \\
			Sensitive attribute & Age  & Gender            & Race    \\
      Social homophily         & 0.9600   & 0.8092    & 0.5361            \\ 
         Average node degree         & 95.79   & 44.48     & 34.04             \\ 
			Graph density  & 47.90 & 22.24             & 17.02   \\
			\hline         
		\end{tabular}
	\end{center}
 \label{data}
\end{table}

\begin{table*}[t]
\Large
\renewcommand\arraystretch{1.3}
\caption{Comparative experiments were conducted on three real-world datasets to evaluate both the validity and fairness of the models. For each metric, $\uparrow$ means larger is better and $\downarrow$ means smaller is better. The dark brown colour is used to highlight the best results for each metric, and the runner-up results are light brown.}
\label{com_exe}
\resizebox{\linewidth}{!}{
\begin{tabular}{lccccccccccc}
\hline
\textbf{Dataset}                     & \textbf{Metrics}       & \textbf{GCN}        & \textbf{GIN}        & \textbf{SAGE}       & \textbf{FairGNN}    & \textbf{NIFTY}      & \textbf{FVGNN}      & \textbf{EDITS}      & \textbf{FairMILE}   & \textbf{FairGB}     & \textbf{EAGNN}       \\
        \hline
\multirow{4}{*}{\textbf{Credit}}     & ACC    $(\uparrow)$         & 73.62±0.06 & 75.30±2.86 & 74.20±0.60 & 75.44±3.28 & 73.80±4.75 & 76.06±4.37 & \cellcolor{brown!60}83.73±0.73 & 80.18±0.27 & \cellcolor{brown!30}80.44±0.12 & 79.02±0.24 \\
                            & F1    $(\uparrow)$          & 81.88±0.06 & 84.56±2.17 & 82.45±0.52 & 81.35±1.83 & 81.21±0.59 & 84.43±4.23 & 76.93±0.89 & 87.16±0.17 & \cellcolor{brown!60}88.35±0.09 & \cellcolor{brown!30}87.96±0.12 \\
                            & $\Delta_{SP} (\downarrow)$ & 12.93±0.26 & 5.14±0.96  & 16.35±2.36 & 10.46±5.69 & 8.09±2.77  & 6.06±3.63  & 7.28±0.49  & \cellcolor{brown!30}1.21±0.39  & 1.29±0.54  & \cellcolor{brown!60}0.41±0.14  \\
                            & $\Delta_{EO} (\downarrow)$ & 10.65±0.18 & 3.79±0.64  & 14.12±2.64 & 9.47±6.10  & 7.41±1.54  & 3.90±3.54  & 5.09±0.78  & 0.84±0.14  & \cellcolor{brown!30}0.75±0.37  & \cellcolor{brown!60}0.48±0.17  \\
                                    \hline
\multirow{4}{*}{\textbf{German}}     & ACC  $(\uparrow)$           & \cellcolor{brown!60}72.45±0.75 & 70.32±1.55 & \cellcolor{brown!30}71.63±1.35 & 70.83±1.66 & 66.24±4.12 & 69.60±1.13 & 65.60±6.81 & 70.08±1.48 & 70.88±0.85 & 70.08±0.16 \\
                            & F1   $(\uparrow)$           & 81.73±2.31 & 81.58±0.56 & 81.08±1.04 & 79.57±2.61 & 78.27±1.25 & 81.33±0.55 & 77.89±6.06 & 80.87±0.94 & \cellcolor{brown!30}82.38±0.34 & \cellcolor{brown!60}82.38±0.04 \\
                            & $\Delta_{SP} (\downarrow)$ & 20.36±5.27 & 6.70±4.92  & 14.33±5.11 & 6.21±2.34  & 8.03±7.19  & 2.50±3.01  & 4.35±4.29  & \cellcolor{brown!30}1.40±0.99  & 3.94±4.30  & \cellcolor{brown!60}0.04±0.09  \\
                            & $\Delta_{EO} (\downarrow)$ & 19.71±5.19 & 5.80±3.32  & 12.53±7.56 & 5.36±2.07  & 4.40±4.18  & 1.26±1.07  & 4.41±3.81  & \cellcolor{brown!30}0.78±0.61  & 1.74±2.57  & \cellcolor{brown!60}0.17±0.34  \\
                            \hline
\multirow{4}{*}{\textbf{Bail}} & ACC  $(\uparrow)$           & 82.49±0.82 & 82.93±0.53 & 87.44±1.34 & 83.56±2.70 & 80.11±5.39 & 87.61±1.30 & 83.15±2.96 & 87.48±0.28 & \cellcolor{brown!60}92.80±0.86 & \cellcolor{brown!30}89.76±0.70 \\
                            & F1  $(\uparrow)$          & 77.52±1.35 & 77.28±0.58 & 81.57±1.19 & 78.37±1.99 & 79.85±3.16 & 82.67±0.87 & 80.42±2.53 & 82.52±0.50 & \cellcolor{brown!60}90.77±0.92 & \cellcolor{brown!30}86.51±0.57 \\
                            & $\Delta_{SP} (\downarrow)$ & 9.31±2.12  & 7.74±1.19  & 8.14±1.08  & 6.88±1.41  & 5.96±2.13  & 3.49±1.74  & 6.57±1.35  & 3.17±0.21  & \cellcolor{brown!30}1.31±1.41  & \cellcolor{brown!60}0.74±0.54  \\
                            & $\Delta_{EO} (\downarrow)$ & 8.59±1.13  & 6.77±0.81  & 7.43±1.75  & 5.77±1.48  & 5.57±1.69  & 2.42±1.29  & 5.61±1.73  & 1.72±0.56  & \cellcolor{brown!30}1.28±0.77  & \cellcolor{brown!60}0.55±0.34 \\
                            \hline
\end{tabular}}
\end{table*}
We employed three well-known datasets, namely the Recidivism,   Credit, and   German datasets~\citep{agarwal2021towards, dai2022learning, wang2022improving, dong2022edits}. The details of these datasets are summarised in Table \ref{data}. 

\begin{itemize}
    \item \textbf{Credit}~\cite{yeh2009comparisons}. Each node in the dataset represents a client, with 13 attributes such as marital status, age, and maximum payment amount. We use age as the sensitive attribute in our experiments.
    \item \textbf{German}~\cite{Dua:2019}. Each node represents a credit card user, and the dataset includes 27 attributes such as employment status, gender, and income. We use gender as the sensitive attribute in our experiments.
    \item \textbf{Bail}~\citep{jordan2015effect}. The nodes in this dataset represent defendants on bail, each with 18 attributes such as type of case, race, and case duration. Race is used as the sensitive attribute in our experiments.
\end{itemize}

The social homophily of each dataset is calculated according to Definition~\ref{socialhomo}. To better analyse the differences between the datasets, we define density as the ratio of the number of edges to the number of nodes. It is worth noting that although the sensitive attributes in the three real-world datasets we have chosen are discrete, our EAGNN method can be directly applied to continuous sensitive attributes. In terms of generality, our method is superior to others.

\subsubsection{Baseline}
In our experiments, we compare the proposed EAGNN method with nine state-of-the-art algorithms. Specifically, these methods can be divided into two categories: (1) \textbf{Vanilla GNNs} and (2) \textbf{Fair GNNs}. The following three methods belong to the category of Vanilla GNNs:
GCN~\cite{kipf2017semisupervised} captures local graph structure features by aggregating information from neighbouring nodes through convolution operations. GIN~\cite{xu2018how} employs a fine-grained feature aggregation mechanism to effectively distinguish nodes across different graphs, enhancing graph isomorphism discrimination, and making it suitable for complex graph structure analysis. SAGE~\cite{hamilton2017inductive} utilises sampling and aggregation strategies, enabling efficient training on large-scale and dynamic graphs while flexibly accommodating changes in node features. 

The following six methods belong to the category of Fair GNNs:
FairGNN~\citep{dai2022learning} addresses bias and discrimination in GNN predictions by leveraging limited sensitive attributes and graph structures. NIFTY~\citep{agarwal2021towards} establishes a novel framework that connects counterfactual fairness with stability in GNNs, facilitating the learning of fair and stable representations. EDITS~\citep{dong2022edits}  creates fairer GNNs from both feature and structural perspectives, mitigating biases present in the input graph. FVGNN~\citep{wang2022improving} targets discriminatory bias by effectively addressing variations in feature correlations during propagation through feature masking strategies. FairMILE~\citep{he2023fairmile} is a multi-level GNN framework designed to learn fair representations while incorporating fairness constraints. FairGB~\cite{li2024rethinking} achieves rebalancing across groups through counterfactual data augmentation and contribution alignment loss.

\subsubsection{Evaluation metrics and implementation details}

In this study, we regard the F1 score (F1) and accuracy (ACC) as the metrics for evaluating the effectiveness of our approach. For the fairness metrics, we use $\Delta_{SP}$ and $\Delta_{EO}$ introduced in Section 2, with smaller values for these fairness metrics indicating fairer model decisions. Following the setup of previous work~\cite{agarwal2021towards, li2024rethinking}, the dataset is divided into three phases: training, validation, and testing.  All FairGNNs use SAGE as the encoder, and the Adam optimisation algorithm is applied across all models. Hyperparameters were tuned in our experiments using a grid search method, and a detailed hyperparameter analysis is presented in Section 5.3.

\subsection{Performance comparison}
\begin{figure*}[ht]
  \centering	
\subfigure[Results for ACC and F1 on Credit ]{\includegraphics[width=0.24\textwidth]{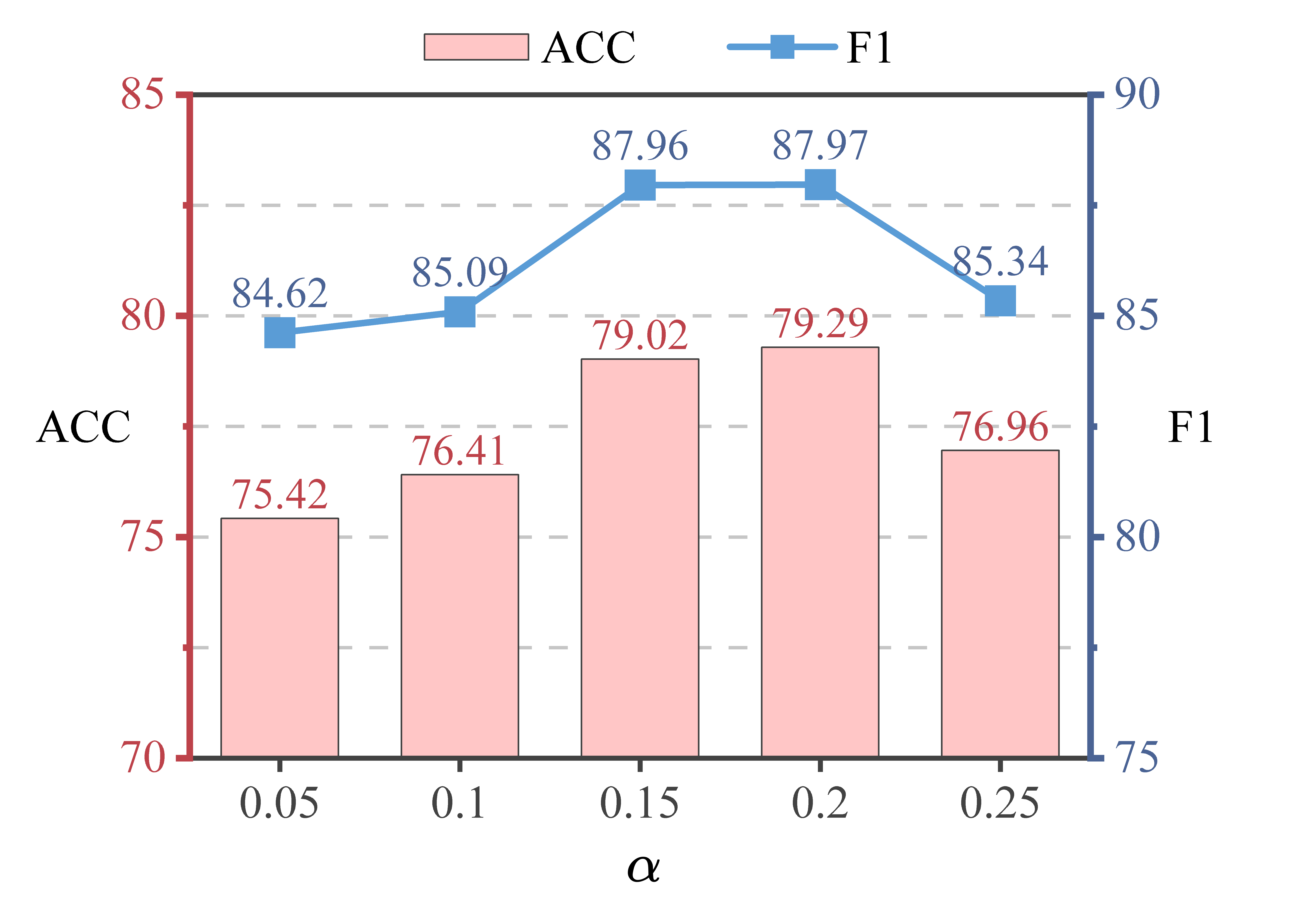} } 		\subfigure[Results for $\Delta_{SP}$ and $\Delta_{EO}$ on Credit]{\includegraphics[width=0.24\textwidth]{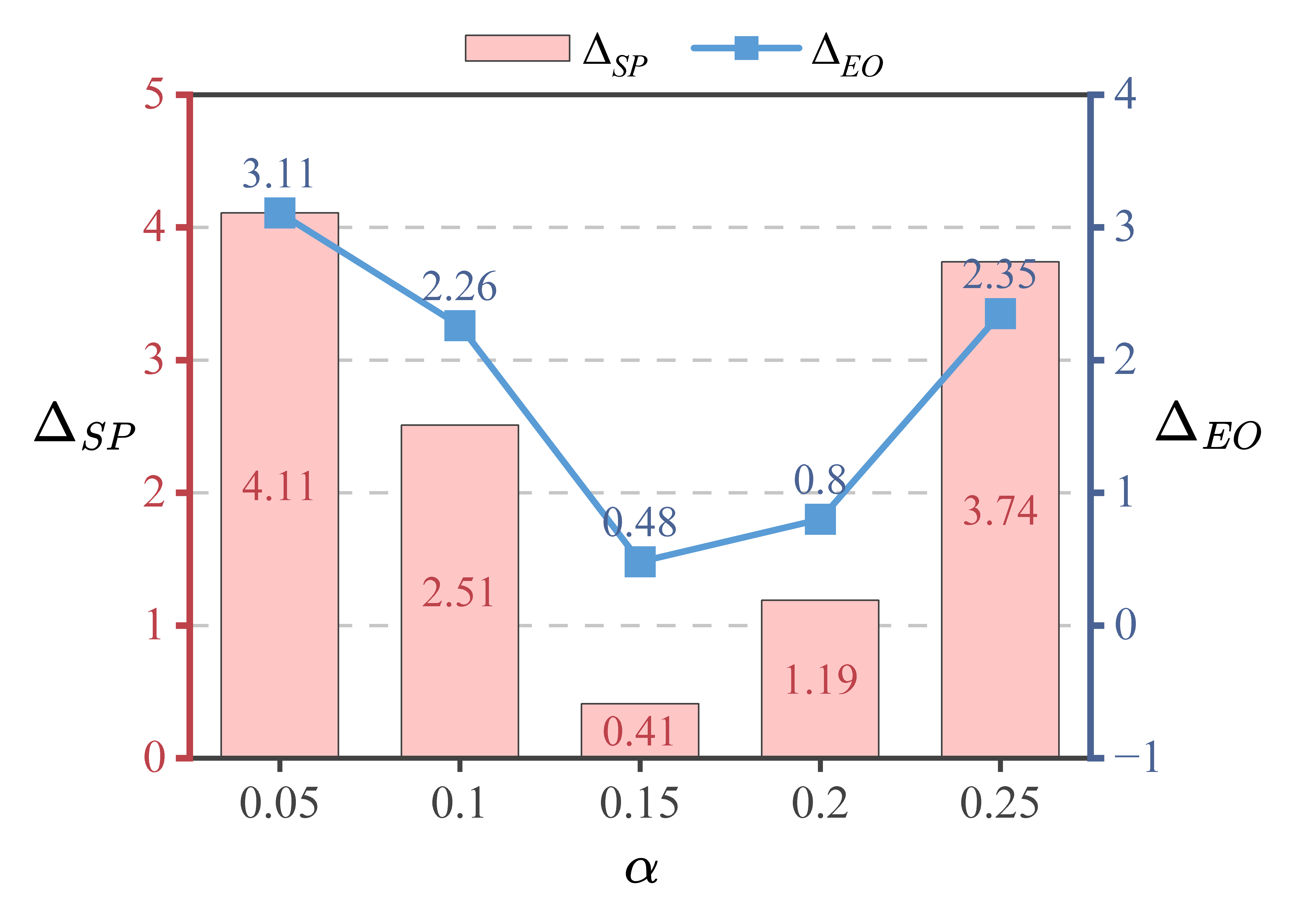} }  
\subfigure[Results for ACC and F1 on German]{\includegraphics[width=0.24\textwidth]{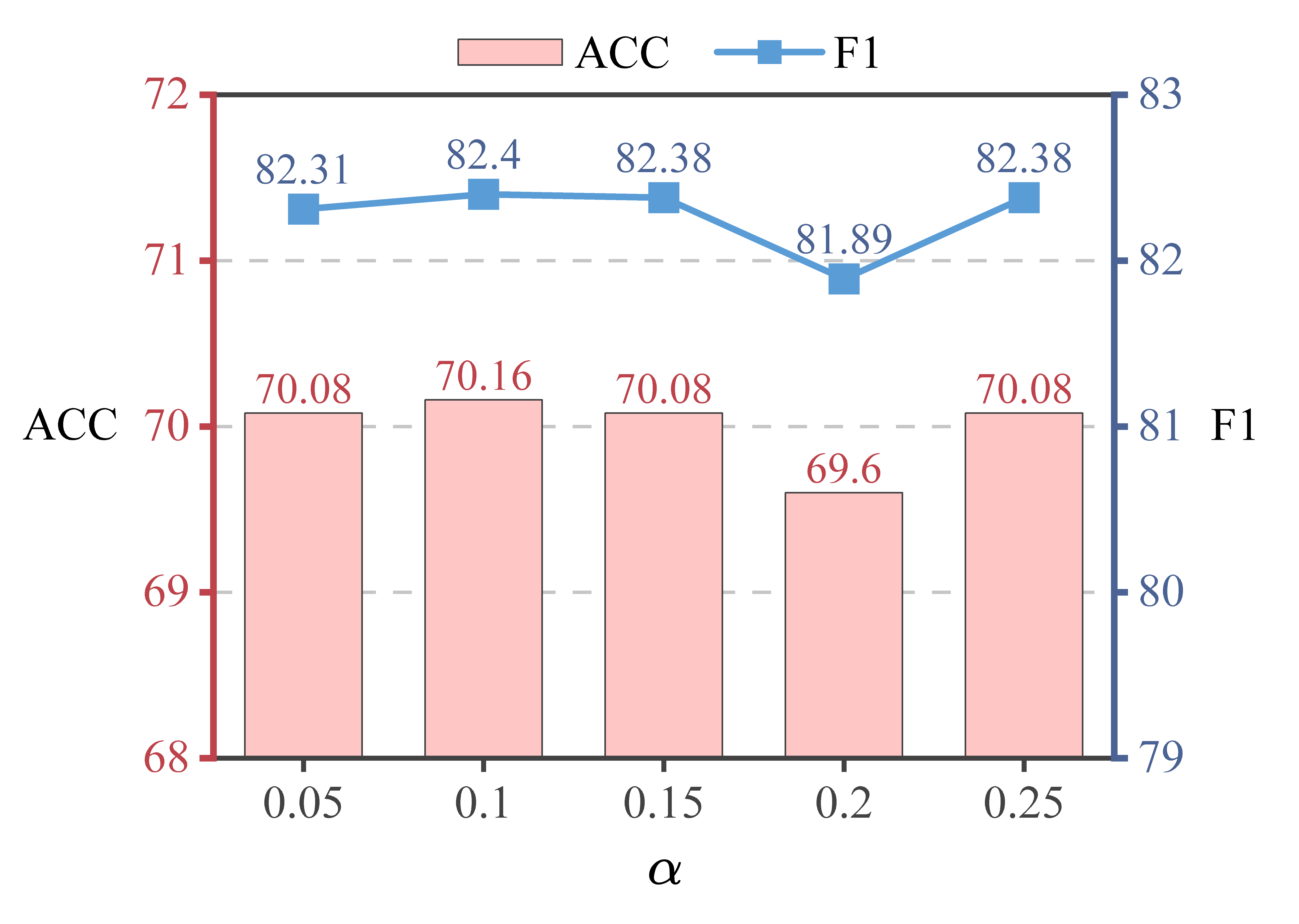} } 		\subfigure[Results for $\Delta_{SP}$ and $\Delta_{EO}$ on German]{\includegraphics[width=0.24\textwidth]{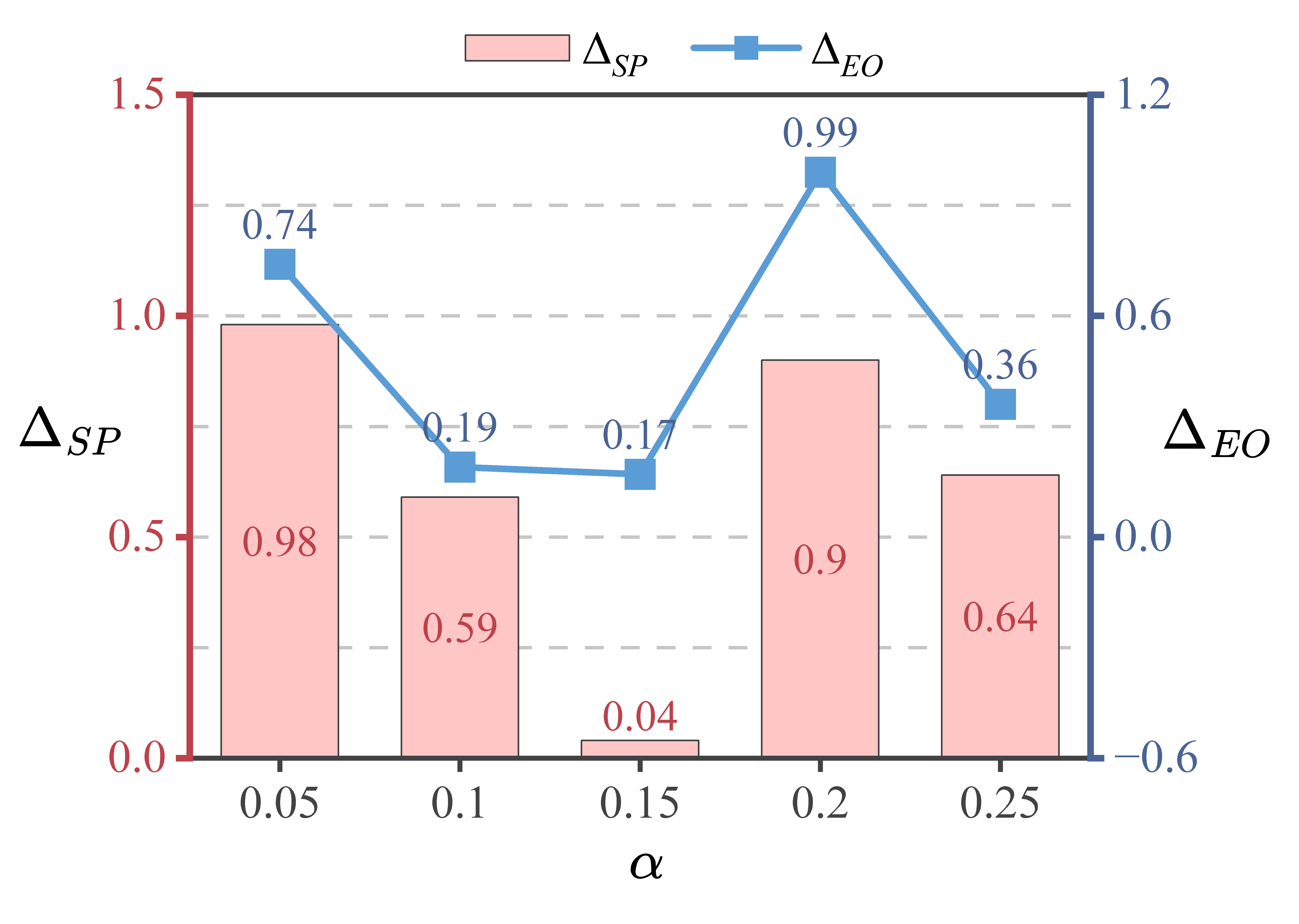} }
	\caption{Sensitivity analysis for the $\mathcal{L}_{suff}$ on Credit and German.}
	\label{sens_a}
\end{figure*}

To gain a comprehensive understanding of EAGNN, we perform node classification tasks on three widely used real-world datasets, comparing EAGNN with other methods. The experimental results are reported in Table~\ref{com_exe}. From  Table~\ref{com_exe}, we have four key observations:
\begin{itemize}
    \item EAGNN achieves satisfactory results in terms of both effectiveness and fairness across all three datasets, often obtaining competitive or even better performance compared to well-designed fairness GNN models. Notably, in some cases, fairness GNN models outperform vanilla GNNs in terms of validity, suggesting that the inherent bias in the dataset reduces the validity of the GNN, making it unreliable. Therefore, it is necessary to debias GNNs not only to improve fairness but also to enhance their overall validity.
   \item EAGNN consistently achieves the best performance in terms of fairness on all datasets. This validates our hypothesis that mitigating social homophily can help GNNs learn fair representations. The three constraints—sufficiency, independence, and separation—effectively prevent spurious correlations between $S$ and $Y$.
   \item On the Bail dataset, we observe that state-of-the-art fairness algorithms tend to outperform vanilla GNNs in both effectiveness and fairness. This is because the edges of Bail dataset is sparse, and the limited graph structure hinders the generalization of GNNs. Fairness GNNs, in their effort to debias, improve overall effectiveness while pursuing fairness.
   \item On the Bail dataset, where social homophily is small, the fairness metrics for all methods are relatively low, further demonstrating the impact of social homophily on GNN fairness. However, GIN’s performance does not stand out on the Credit and German datasets, where social homophily is higher. In datasets with high social homophily, the graph is more densely connected, and nodes may share very similar features and connectivity patterns. GIN, which aggregates neighbourhood information to determine node importance, may overlook differences between groups, thus achieving better fairness.
\end{itemize}

\subsection{Ablation study}

\begin{figure*}[hbt]
  \centering	
\subfigure[Results for ACC on Credit]{\includegraphics[width=0.24\textwidth]{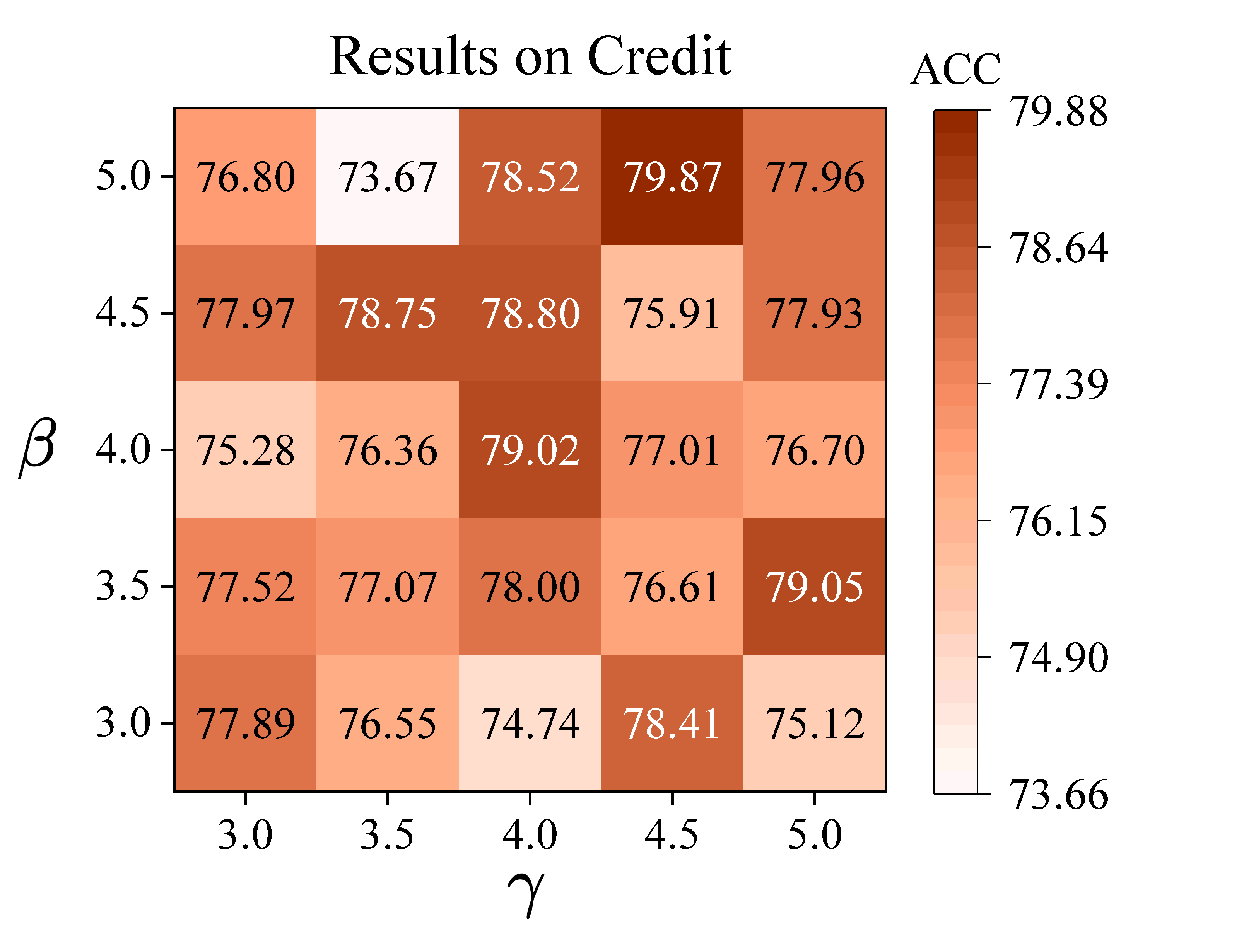} } 		\subfigure[Results for F1 on Credit]{\includegraphics[width=0.24\textwidth]{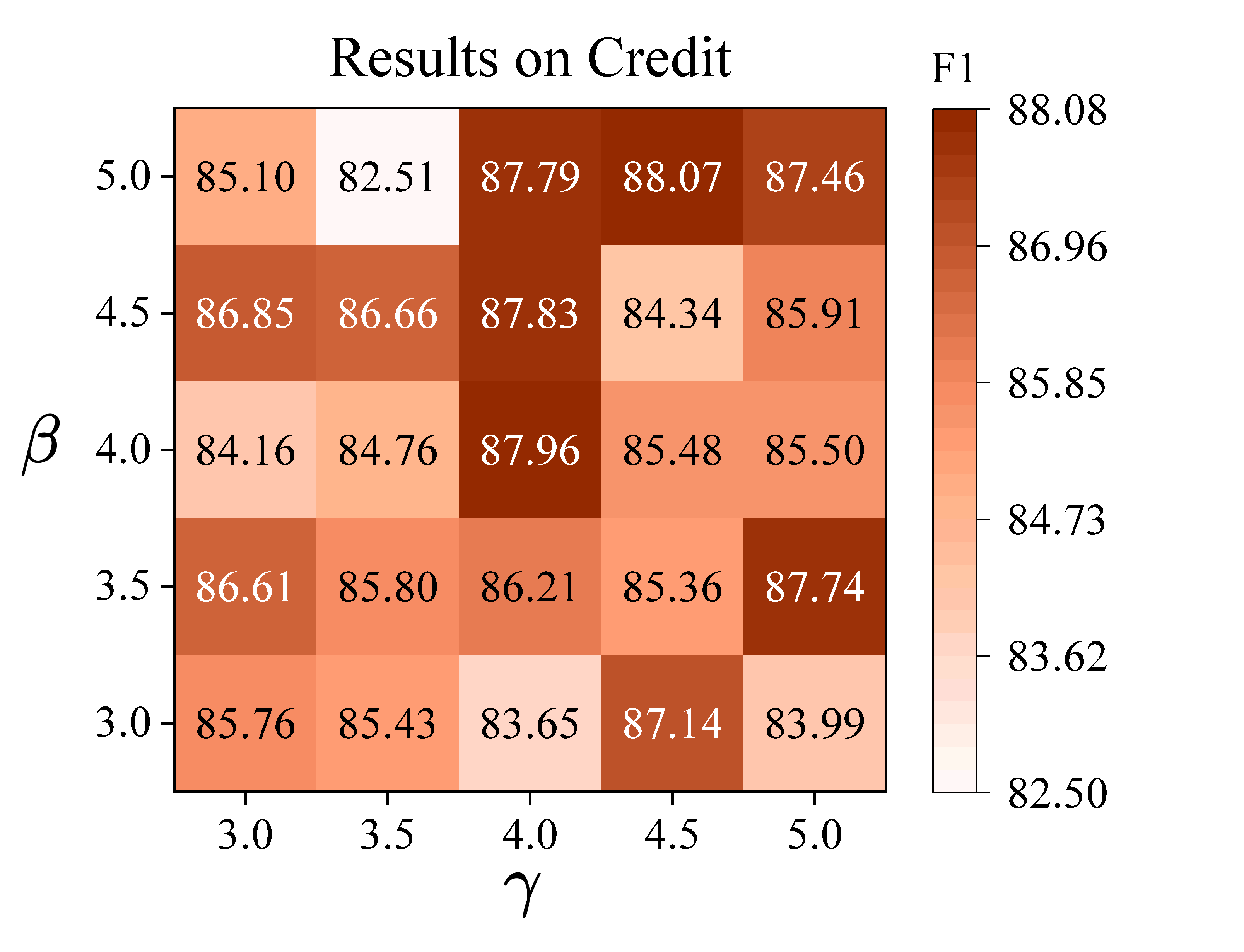} }
\subfigure[Results for $\Delta_{SP}$ on Credit]{\includegraphics[width=0.24\textwidth]{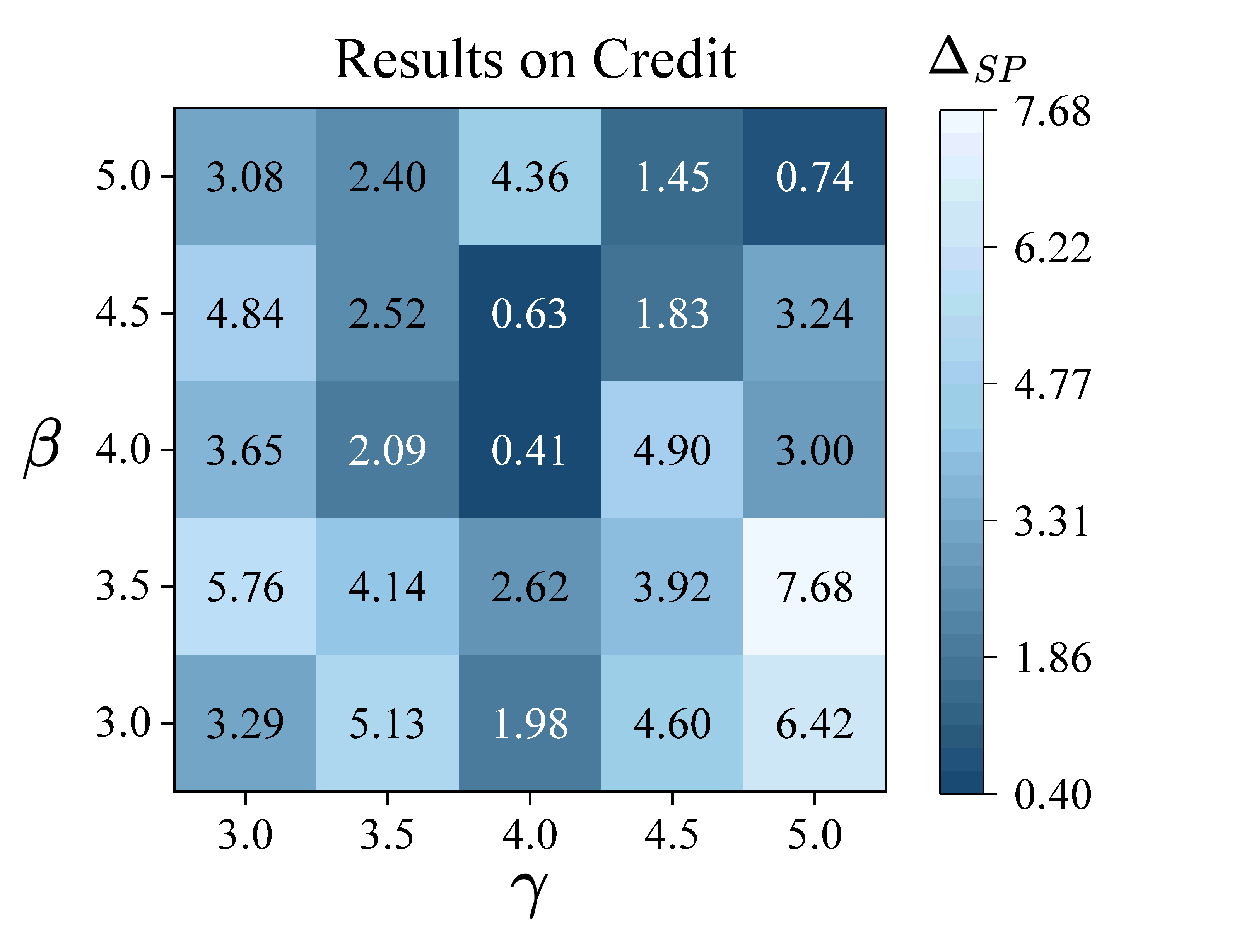} }
\subfigure[Results for $\Delta_{EO}$ on Credit]{\includegraphics[width=0.24\textwidth]{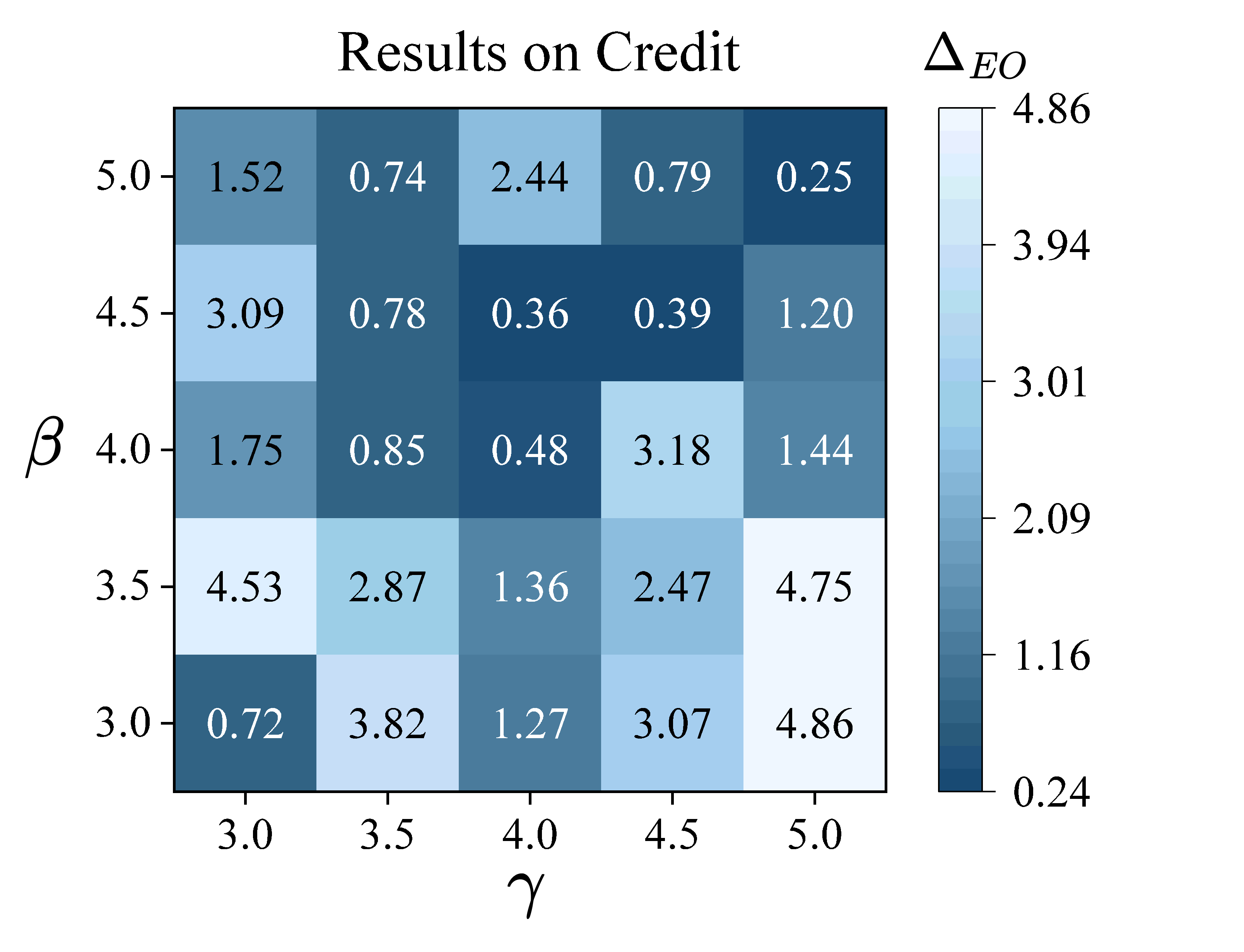} }
\subfigure[Results for ACC on German]{\includegraphics[width=0.24\textwidth]{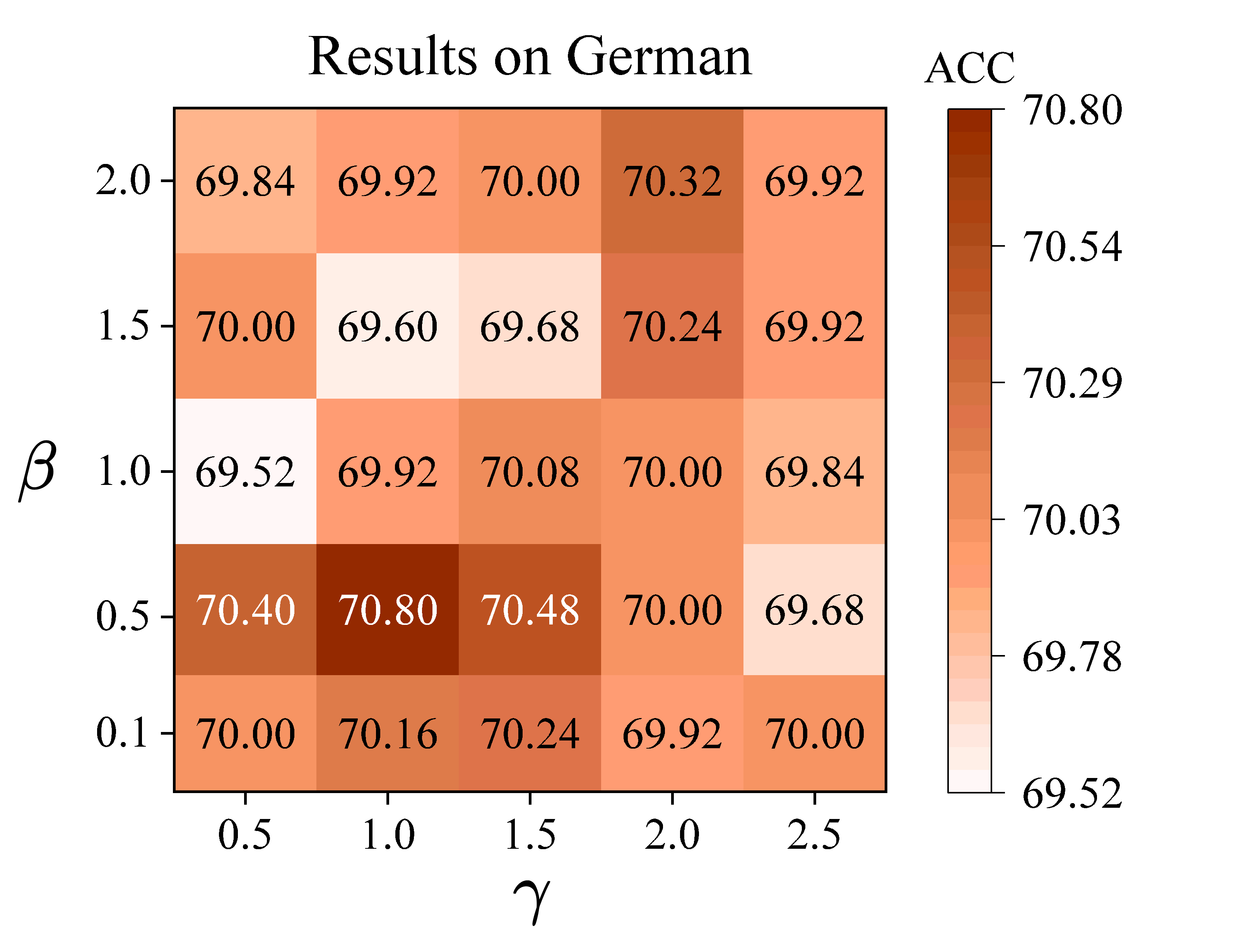} } 		\subfigure[Results for F1 on German]{\includegraphics[width=0.24\textwidth]{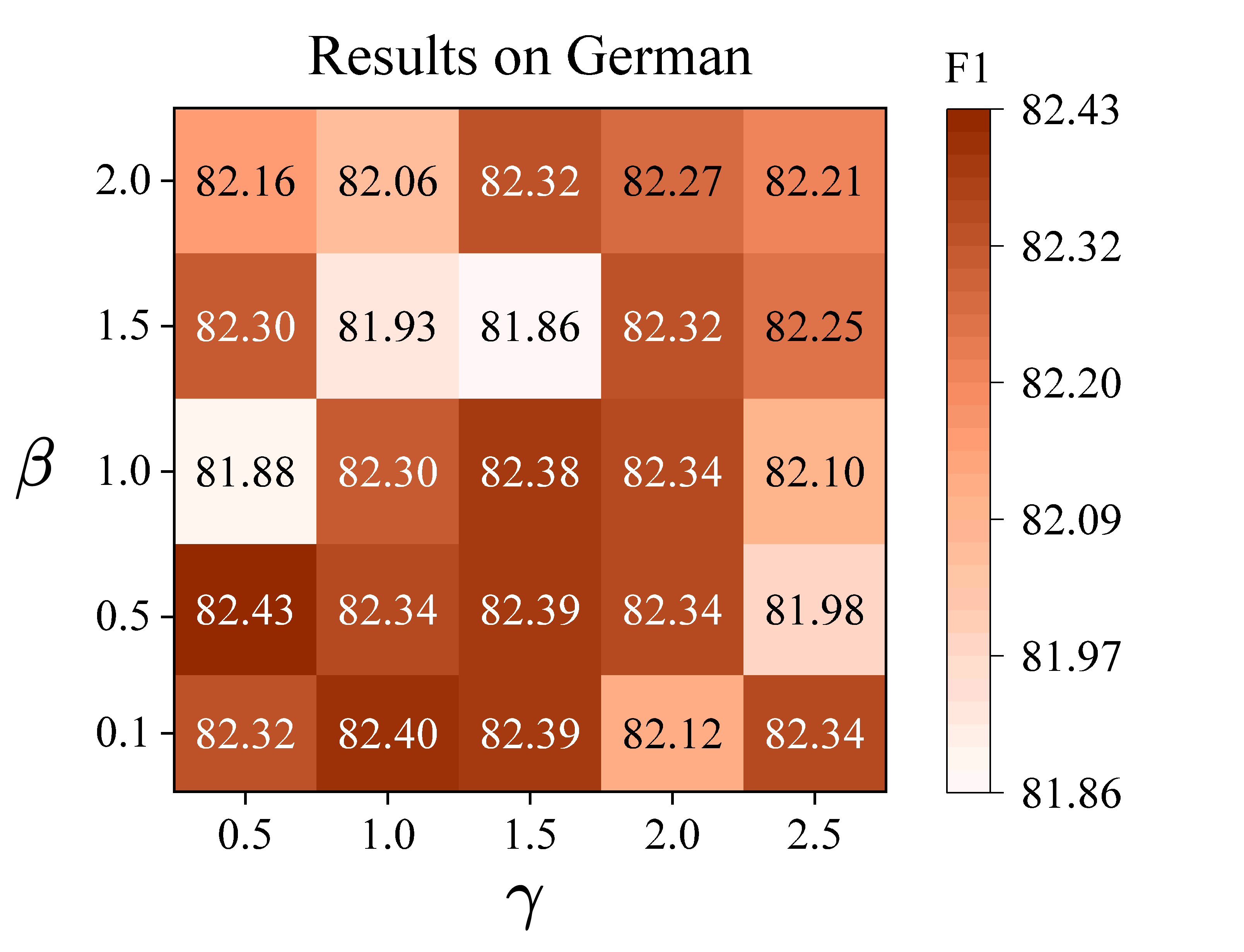} }
\subfigure[Results for $\Delta_{SP}$ on German]{\includegraphics[width=0.24\textwidth]{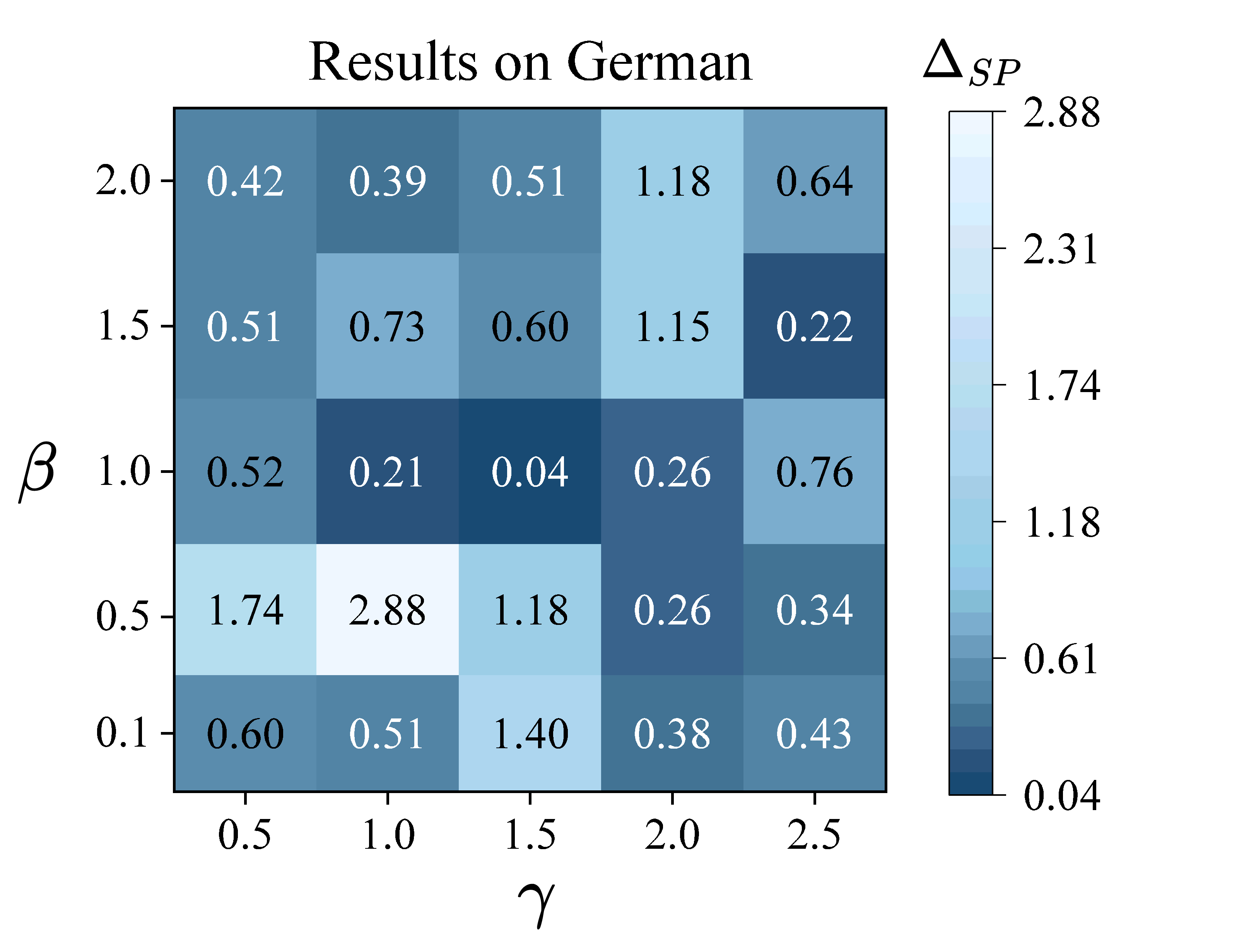} }
\subfigure[Results for $\Delta_{EO}$ on German]{\includegraphics[width=0.24\textwidth]{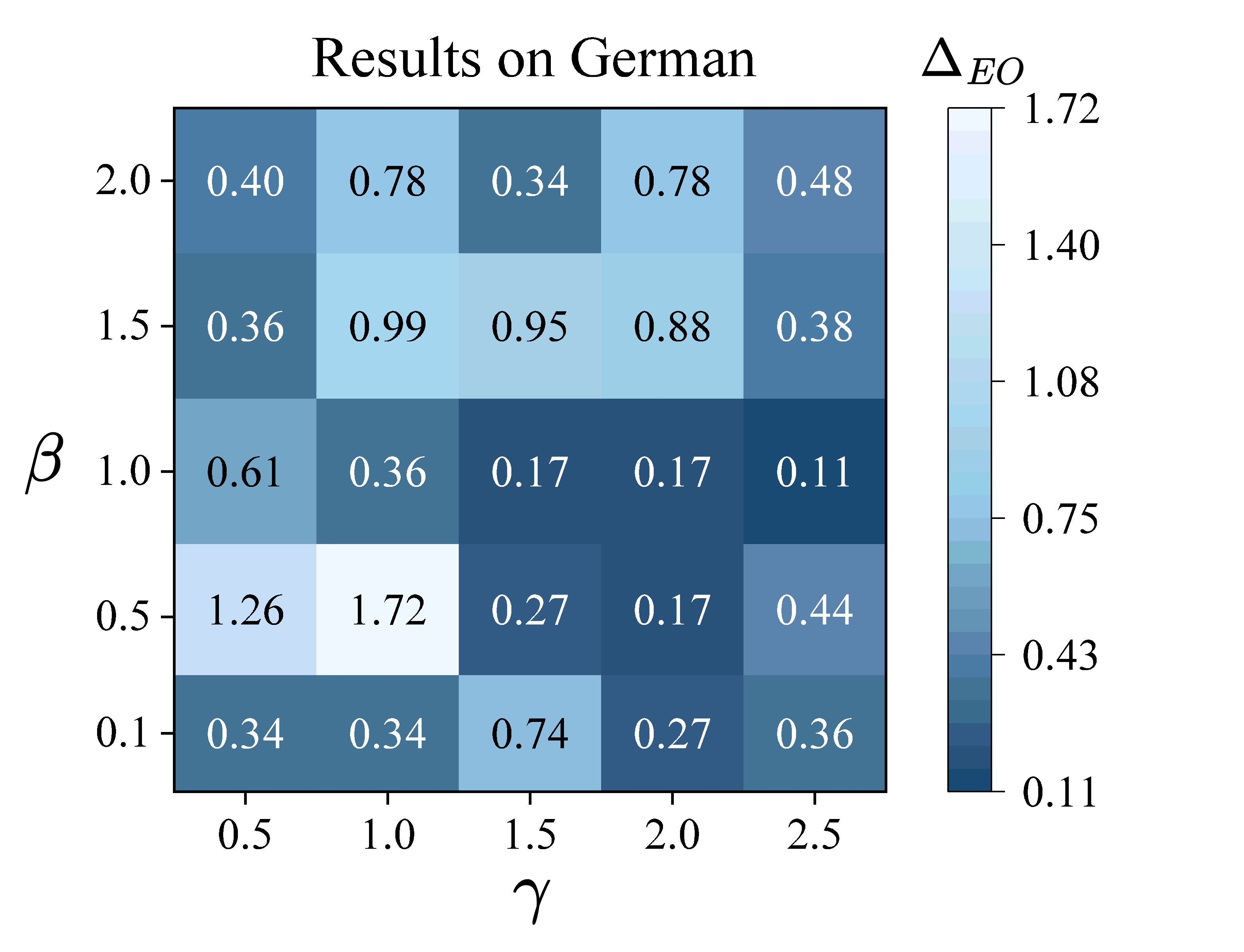} }
\subfigure[Results for ACC on Bail]{\includegraphics[width=0.24\textwidth]{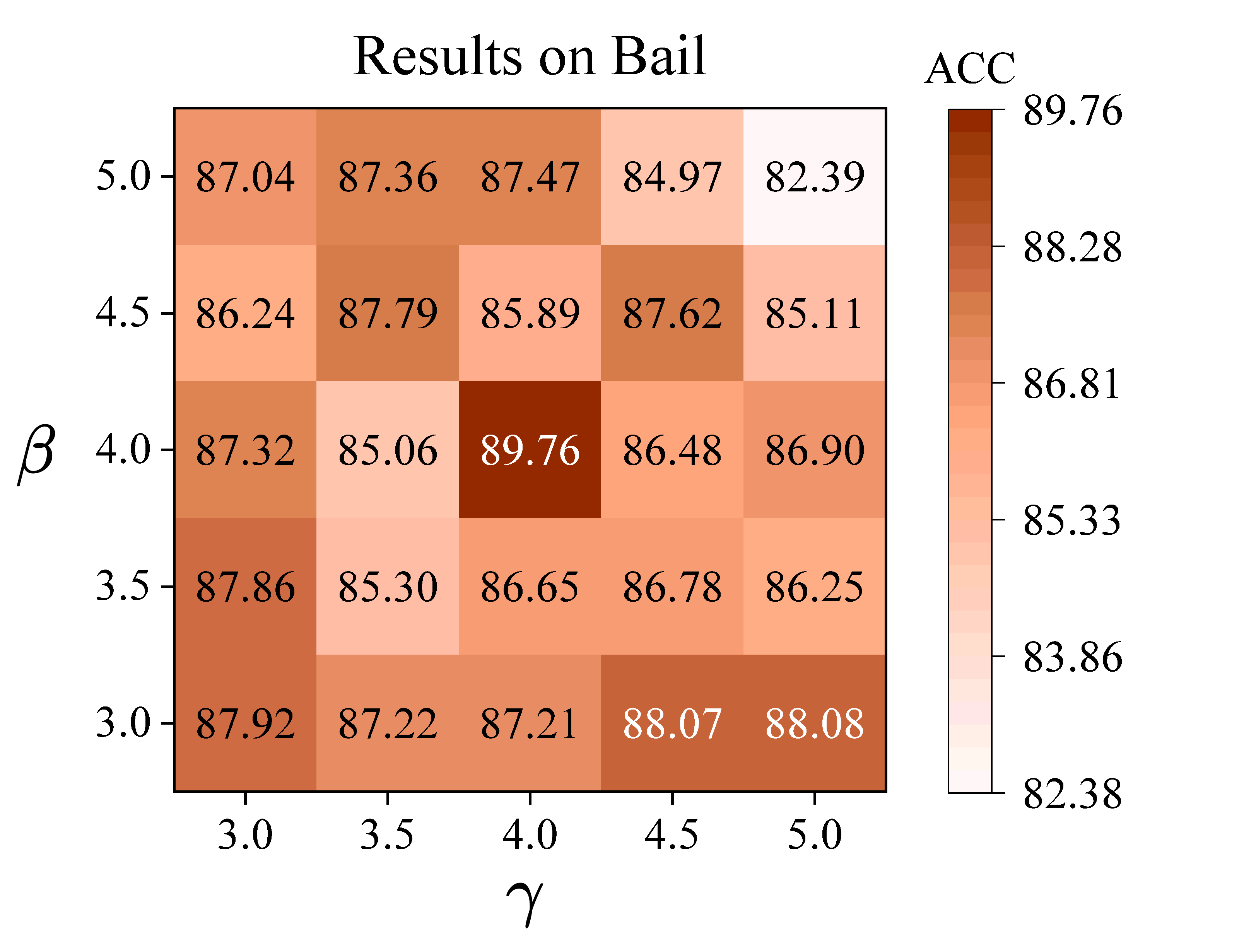} } 		\subfigure[Results for F1 on Bail]{\includegraphics[width=0.24\textwidth]{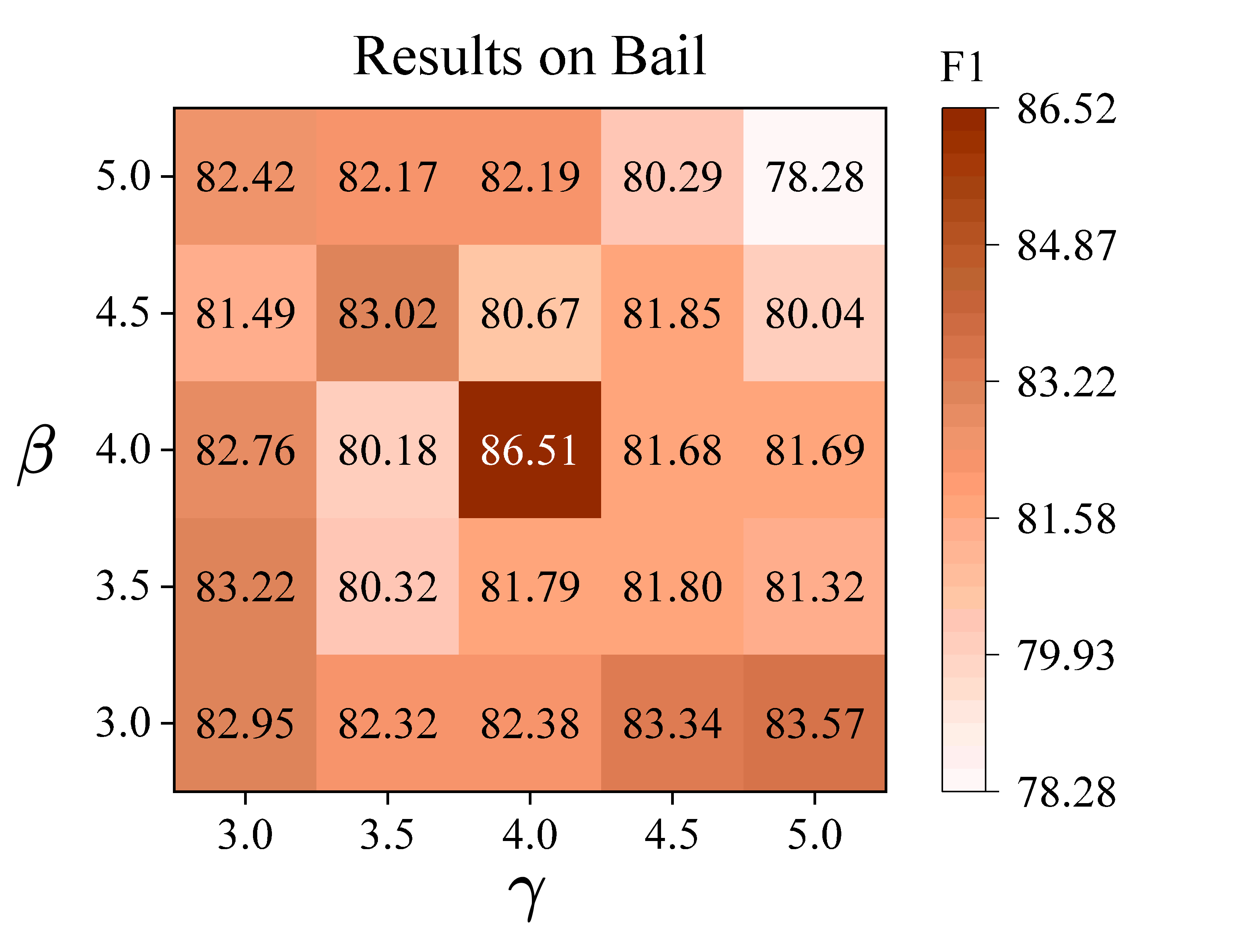} }
\subfigure[Results for $\Delta_{SP}$ on Bail]{\includegraphics[width=0.24\textwidth]{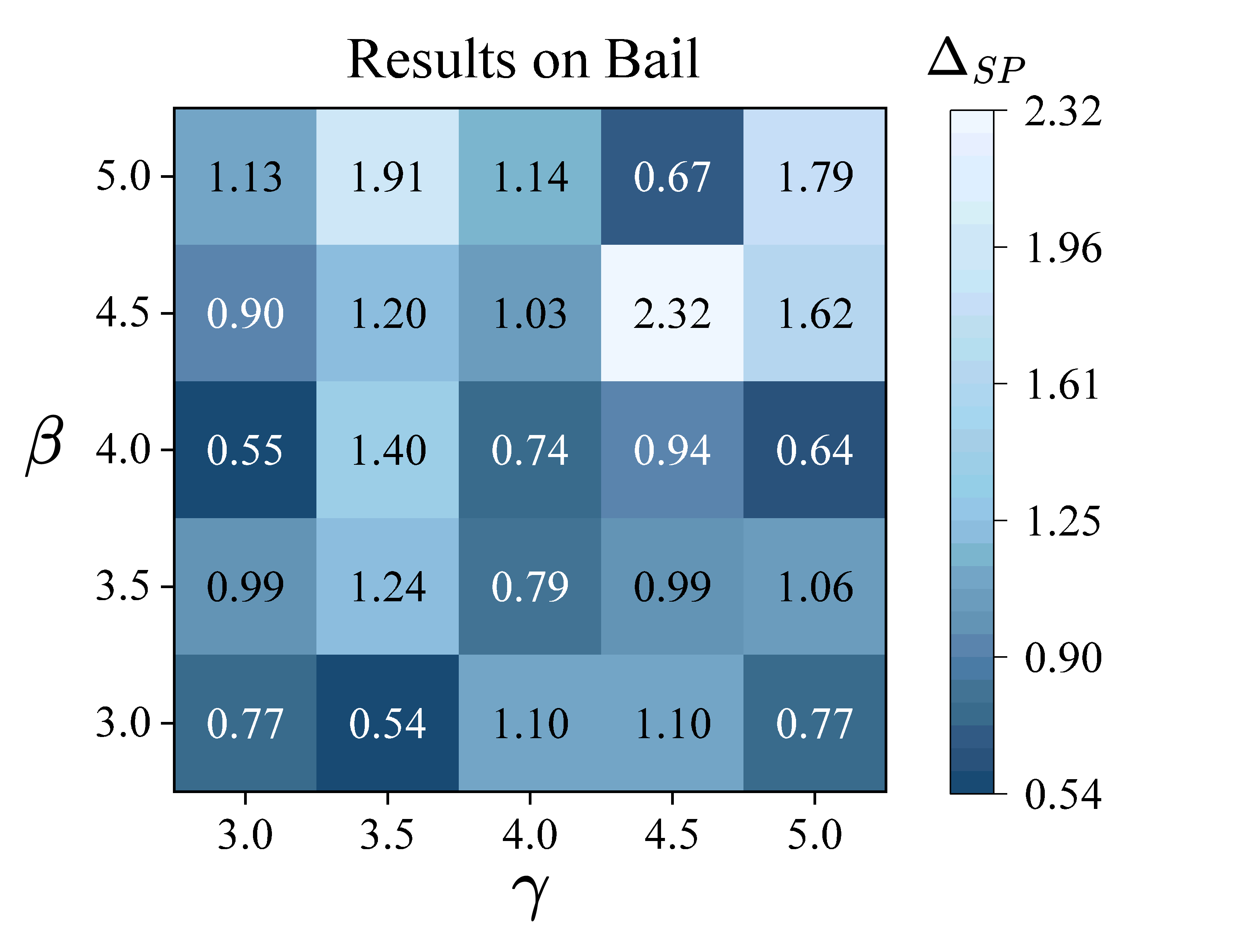} }
\subfigure[Results for $\Delta_{EO}$ on Bail]{\includegraphics[width=0.24\textwidth]{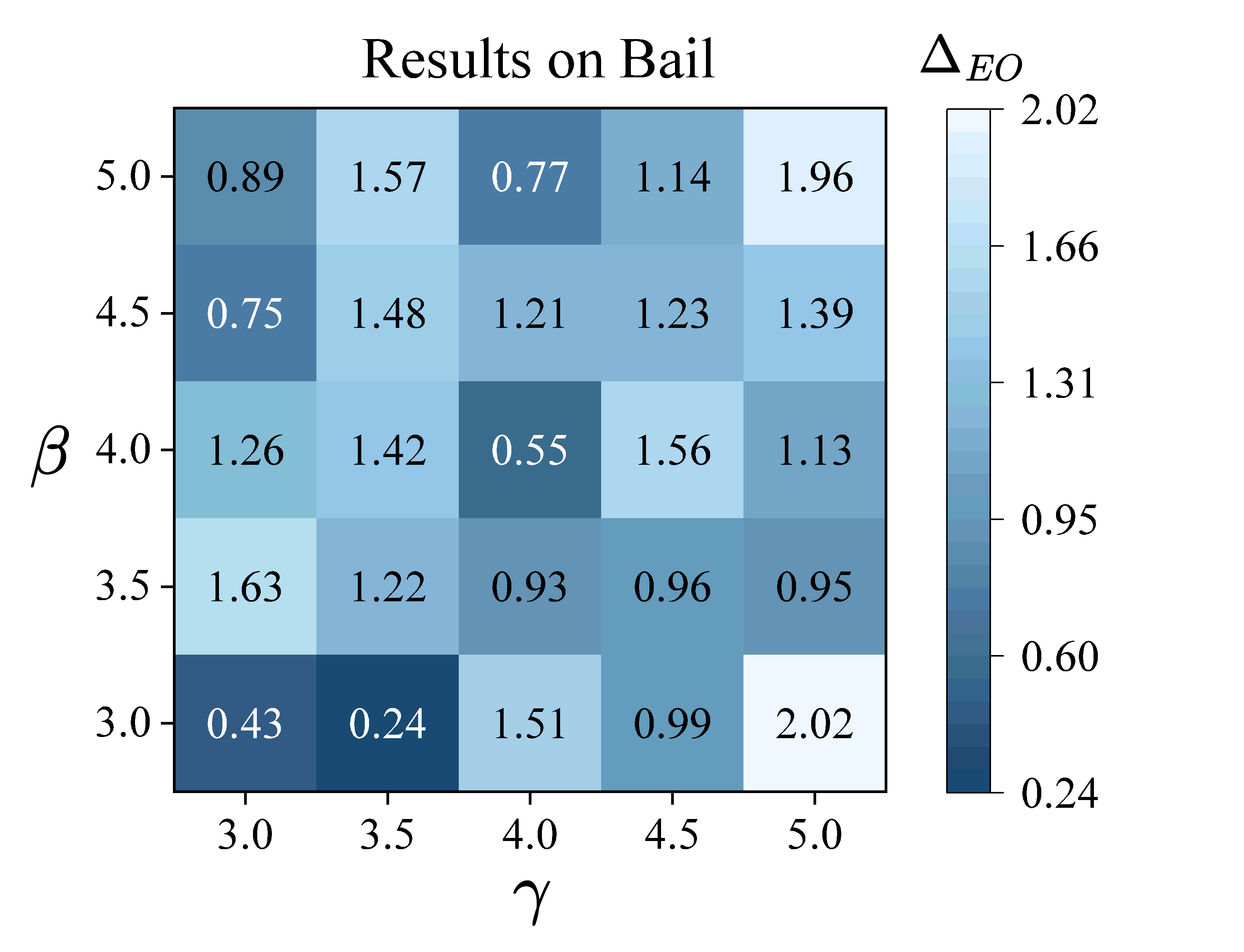} }
	\caption{Sensitivity analysis for the $\mathcal{L}_{in}$  and $\mathcal{L}_{se}$ on three real-world datasets.}
	\label{sens_b}
\end{figure*}

\begin{figure}[h]
  \centering	
\subfigure[ACC and F1 on Bail]{\includegraphics[width=0.23\textwidth]{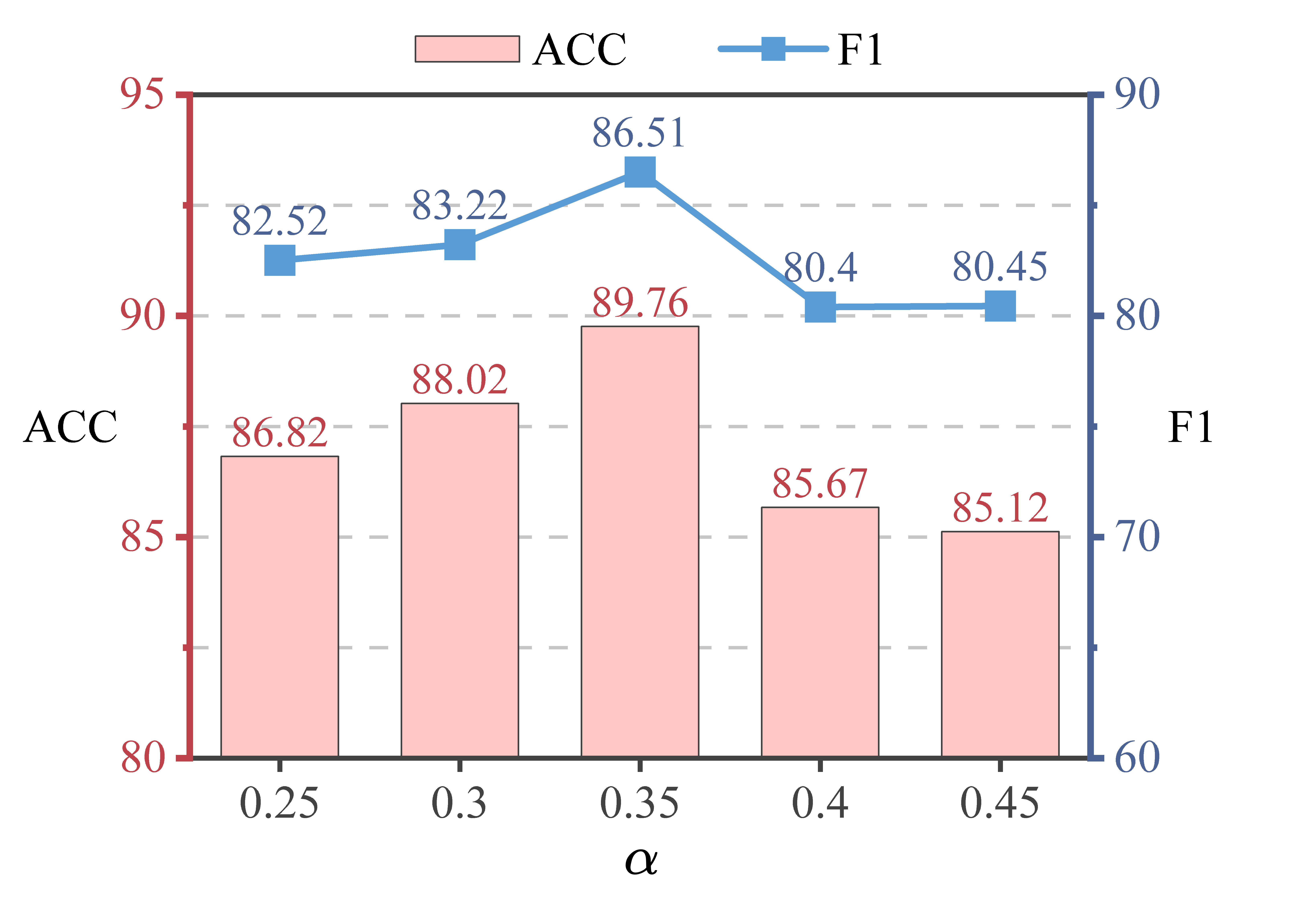} }
\subfigure[$\Delta_{SP}$ and $\Delta_{EO}$ on Bail]{\includegraphics[width=0.23\textwidth]{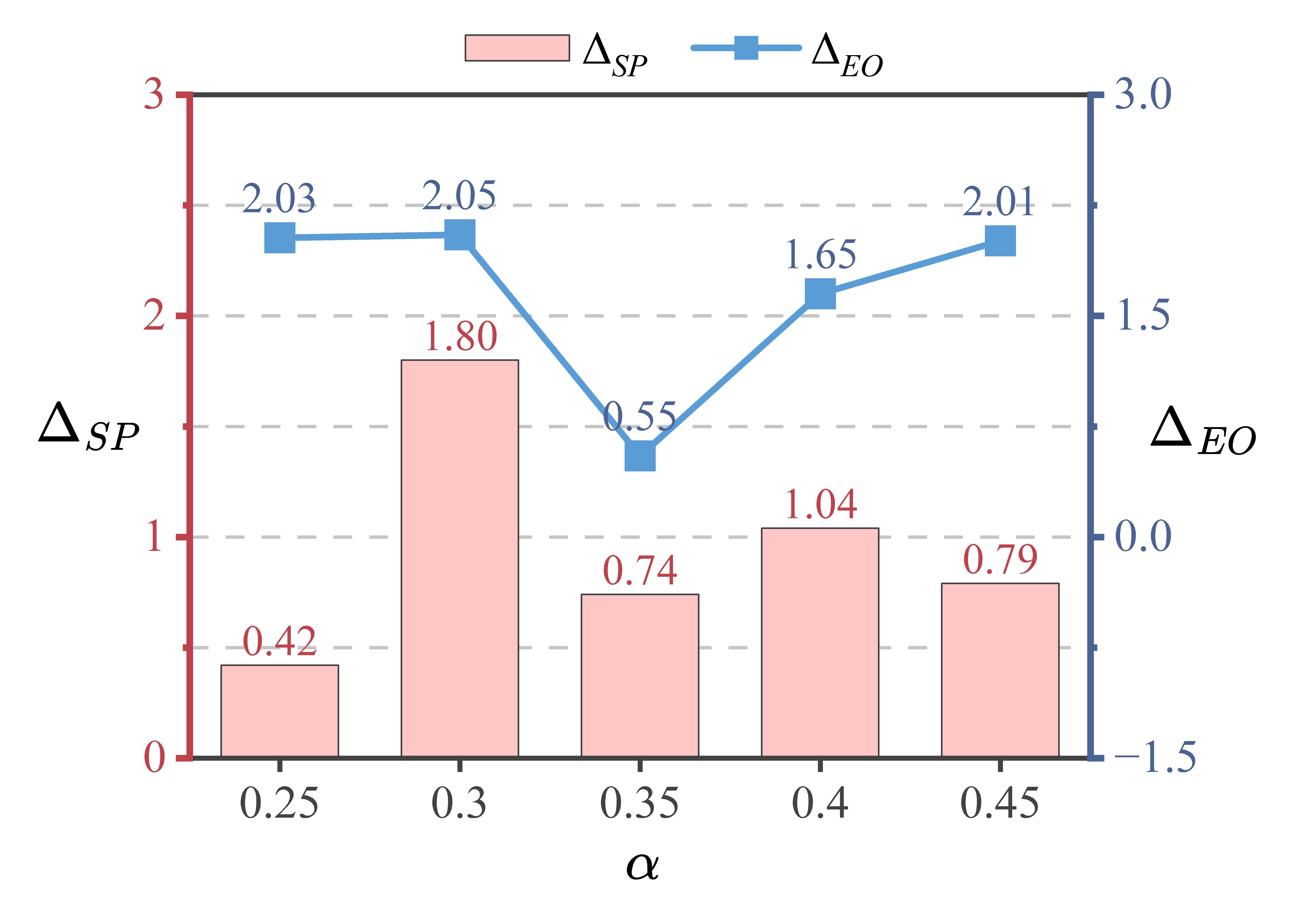} }
	\caption{Sensitivity analysis for $\mathcal{L}_{suff}$ on Bail.}
	\label{sens_bail}
\end{figure}

\begin{table}[ht]
\renewcommand\arraystretch{1.3}
\caption{Ablation study results. For each metric, $\uparrow$ means larger is better and $\downarrow$ means smaller is better.}
\label{abl}
\resizebox{\linewidth}{!}{
\begin{tabular}{lccccc}
\hline
\textbf{Dataset }                    & \textbf{Metrics}       & \textbf{w/o} $\mathcal{L}_{suff}$ & \textbf{w/o} $\mathcal{L}_{in}$  & \textbf{w/o} $\mathcal{L}_{se}$  & \textbf{EAGNN}       \\
\hline
\multirow{4}{*}{\textbf{Credit}}     & ACC  $(\uparrow)$         & 71.94±5.22  & 77.27±0.56 & 75.21±2.58 & 79.02±0.24 \\
                            & F1   $(\uparrow)$         & 80.84±5.43  & 85.64±1.06 & 83.96±3.00 & 87.96±0.12 \\
                            & $\Delta_{SP} (\downarrow)$ & 2.00±1.18   & 1.36±1.43  & 1.39±1.12  & 0.41±0.14  \\
                            & $\Delta_{EO} (\downarrow)$ & 0.76±0.49   & 0.76±0.80  & 0.81±0.60  & 0.48±0.17  \\
                            \hline
\multirow{4}{*}{\textbf{German}}     & ACC $(\uparrow)$          & 70.48±0.59  & 70.00±0.04 & 69.92±0.16 & 70.08±0.16 \\
                            & F1  $(\uparrow)$          & 82.35±0.11  & 82.27±0.13 & 82.16±0.15 & 82.38±0.04 \\
                            & $\Delta_{SP} (\downarrow)$ & 2.80±3.43   & 0.42±0.66  & 1.23±0.77  & 0.04±0.09  \\
                            & $\Delta_{EO} (\downarrow)$ & 1.32±2.49   & 0.74±0.90  & 1.64±1.06  & 0.17±0.34  \\
                            \hline
\multirow{4}{*}{\textbf{Bail}} & ACC $(\uparrow)$          & 86.67±0.52  & 87.02±0.51 & 86.06±1.53 & 89.76±0.70 \\
                            & F1   $(\uparrow)$         & 82.01±0.42  & 82.50±0.54 & 81.15±1.27 & 86.51±0.57 \\
                            & $\Delta_{SP} (\downarrow)$ & 0.81±0.55   & 1.04±0.36  & 0.79±0.30  & 0.74±0.54  \\
                            & $\Delta_{EO} (\downarrow)$ & 0.63±0.19   & 1.77±0.69  & 0.65±0.52  & 0.55±0.34 \\
                            \hline
\end{tabular}}
\end{table}

To verify the necessity of each component in EAGNN, we constructed three variants of EAGNN by removing the sufficiency constraint (\textbf{w/o} $\mathcal{L}_{suff}$), the independence constraint (\textbf{w/o} $\mathcal{L}_{in}$), and the separation constraint (\textbf{w/o} $\mathcal{L}_{se}$). From the experimental results in Table~\ref{abl}, we have three observations about EAGNN: 
\begin{itemize}
    \item Regardless of which constraint is removed, the fairness of EAGNN decreases, demonstrating the necessity of each module. EAGNN relies on the interplay of the three constraints to prevent spurious associations between $S$ and $Y$ by mitigating social homophily.
   \item The fairness metrics of the model decrease when $\mathcal{L}_{suff}$ is removed. This is because $\mathcal{L}_{suff}$ ensures that nodes belonging to different groups, but with similar attributes, are sufficiently trained, which helps to highlight their differences. Compared to $\Delta_{EO}$, $\Delta_{SP}$ focuses more on the correlation between predictions and groups, leading to a more significant deterioration in SP metrics.
    \item When either $\mathcal{L}_{in}$ or $\mathcal{L}_{se}$ is removed, both fairness metrics decrease, indicating that these two constraints interact with each other. However, for the Credit and German datasets, which exhibit high social homophily, the fairness metrics decrease more significantly when $\mathcal{L}_{se}$ is removed. This is because, in graphs with high social homophily, members within the same group may share many attributes that are, in reality, influenced by the sensitive attribute~\cite{martinez2020sars, chen2022impact}. Thus using GNN to learn representations on graphs with very high social homophily, it is very easy for the model to establish spurious correlations between model predictions and sensitive attributes, even given the node labels. In this case, $\mathcal{L}_{se}$ is more important than $\mathcal{L}_{in}$ for fair representation learning.
\end{itemize}

\subsection{Hyperparameter sensitive analysis}
Moreover, we conduct experiments to analyse the hyperparameters of the three constraints. We first control the independence constraint weight $\beta$ and the separation constraint weight $\gamma$ unchanged, while varying the sufficiency constraint weight $\alpha$ to analyse its impact. The results on the Credit and German datasets, which have high social homophily, are shown in Figure~\ref{sens_a}, and the results for the Bail dataset, which has low social homophily, are presented in Figure~\ref{sens_bail}.

As observed in Figure~\ref{sens_a}, optimal validity and fairness are achieved when the weight of $\mathcal{L}_{suff}$ is set to 0.15 for the Credit and German datasets, whereas for the Bail dataset, it needs to be set to 0.35. This is because low social homophily implies that individuals from different groups may differ significantly in many characteristics. In this case, increasing the sufficiency weight helps the model better learn and understand the features of non-sensitive attributes, reducing misclassification and bias toward these nodes. The sufficiency constraint encourages the model to learn shared features, even when the sensitive attributes differ, resulting in more accurate predictions. Additionally, we observe that when the sufficiency constraint is properly balanced, both fairness and model effectiveness improve. This demonstrates that the sufficiency constraint not only promotes fairness but also enhances classification accuracy by enabling sufficient learning across group boundaries.

For the experiments balancing the independence constraint weight $\beta$ and the separation constraint weight $\gamma$, the results are displayed in Figure~\ref{sens_b}. As shown in Figure~\ref{sens_b}, the model tends to achieve better fairness when $\beta$ and $\gamma$ are set to the same value. Theoretically, both the independence and separation constraints align with fairness principles, and assigning them equal weight reflects a balanced respect for these principles. Overemphasising one constraint could cause the model to overlook another important fairness consideration. Treating both constraints equally helps avoid sacrificing one fairness requirement in favour of another. As observed in Figure~\ref{sens_b}, advanced effectiveness and optimal fairness are obtained when the weight of $\mathcal{L}_{suff}$ is set to 0.15. This suggests that sufficiency not only contributes to fairness but also leads to accurate classification through adequate learning of cross-group nodes.

\section{Related work}
In this section, we review related work on fairness in GNNs and data augmentation, which are most relevant to our EAGNN method.

\subsection{Fairness in GNNs}


There has been a wide variety of work attempting to improve the fairness of GNNs. Fairwalk~\cite{rahman2019fairwalk} and Crosswalk~\cite{khajehnejad2022crosswalk} cross group boundaries by selecting each set of neighbouring nodes with probabilistic dropping or biased random walks. EDITS~\cite{dong2022edits} de-configures attribute and structural information to enhance the fairness of the model. FairGNN~\cite{dai2022learning} enables the model to produce fair outputs through adversarial training with min-max objectives. Subsequent approaches aid adversarial learning through various augmentation methods. NIFTY~\cite{agarwal2021towards} designs a representation learning strategy for GNNs that both reduces bias and improves robustness by introducing a new objective function that takes both fairness and stability into account, and by combining it with a hierarchical weight normalisation method that uses Lipschitz's constant. FVGNN~\citep{wang2022improving} targets discriminatory bias by effectively addressing variations in feature correlations during propagation through feature masking strategies.  FairMILE~\citep{he2023fairmile} proposes a multilevel framework that fully integrates existing graph embedding methods.  FDGNN~\cite{zhang2024disentangled} achieves disentanglement based on contrastive learning on node representations. FairGB~\cite{li2024rethinking} achieves rebalancing by interpolating to form new samples. 

However, these methods do not take into account the effect of social homophily, while state-of-the-art methods require complex designs. In this paper, we analyse the effect of social homophily on the fairness of GNNs and achieve simple and effective learning of fair representations through the three aspects of sufficiency, independence, and separation.

\subsection{Data Augmentation in GNNs}
GNN belongs to data-driven deep neural networks, which makes its training results dependent on the quality of data. Some researchers have proposed improving the training results of the model through data augmentation, which can be specifically classified into two categories. (1) The first involves artificially introducing perturbations to the training graph to generate novel training samples, thereby amplifying the dataset and bolstering the model's capacity for generalization across varied graph topologies, a process commonly referred to as data augmentation. Specifically, it includes 1) subgraph sampling~\cite{sun2021sugar, zhang2024noise}, which induces subgraphs by randomly selecting nodes and their neighbours from the original graph; 2) edge modification~\cite{yang2022supervised, kong2022robust}, which randomly removes or adds edges to the graph with a certain probability; and 3) feature masking~\cite{hou2022graphmae, feng2022adversarial}, which partially masks node features. These approaches together enhance the generalisation of the model, creating new training examples while retaining the core topology and inherent patterns of the original graph. (2) Structural or category imbalance is ameliorated by rebalancing ideas to avoid model bias. 
BLC~\cite{zhang2024bayesian} devises strategies to enhance the long tail for the imbalance problem in the structure. GRAPHENS~\cite{park2022graphens} discovers the phenomenon of neighbour memory in the classification of imbalanced nodes and synthesizes self-networks to generate a few nodes based on similarity. GraphSHA~\cite{li2023graphsha} synthesizes only harder training samples and generates connected edges from subgraphs to stop messages from propagating from a few nodes to neighbouring classes. IA-FSNC~\cite{wu2024graph} achieves effective node classification through support augmentation and shot augmentation. HyperIMBA~\cite{fu2023hyperbolic} improves structural imbalance from the perspective of hyperbolic geometry.

However, it is important to note that our EAGNN method differs from previous data augmentation approaches. While EAGNN randomly generates sensitive attribute values for each node, it does not modify the data itself, node representations for both training and prediction are based on the original data. We theoretically demonstrate that EAGNN can achieve group fairness on independence and separation.

\section{Conclusion}
In this paper, we provide a novel perspective on addressing the fairness issue in GNNs. We identify social homophily as a significant factor contributing to unfairness in GNNs. We demonstrate that the message-passing mechanism of GNNs tends to reinforce group-based biases due to social homophily, resulting in spurious associations between sensitive attributes and model predictions. To mitigate these effects, we propose the EAGNN method, which enhances fairness through constraints on three key aspects: sufficiency, independence, and separation. Our theoretical analysis confirms that these constraints effectively reduce bias and promote group fairness in GNN predictions. Additionally, the EAGNN method is broadly applicable to various fairness scenarios, regardless of whether sensitive attributes are continuous or discrete. Extensive experiments on three real-world datasets with varying degrees of social homophily demonstrate that our EAGNN achieves the state-of-the-art performance across two fairness metrics and offers competitive effectiveness.


\bibliographystyle{ACM-Reference-Format}
\bibliography{sample}

\appendix
\setcounter{theorem}{0}
\setcounter{equation}{0}

\section{Theoretical proof}

\subsection{Social homophily effects}
\label{proof1}
\begin{theorem}
Let $\mathcal{G}$ be a graph defined by ${\mathcal{V}, \mathcal{E}}$. Each node  $v_i$  in  $\mathcal{G}$ is characterized by a feature vector  $\mathbf{x}_i \in \mathbb{R}^l$ and a sensitive attribute $s_i$. For any node $ v_i \in \mathcal{V} $ of group $b$, the expectation of the pre-activation output of a single GCN operation is given by:

\begin{equation}
    \mathbb{E} \left[ \mathbf{h_i} \right] = \mathbf{W} \left( \mathbb{E}_{b \sim \mathcal{D}_{s_{i}}, \mathbf{x} \sim \mathcal{F}_{b}} \left[ x \right] \right), 
\end{equation}
where $\mathbf{W}$ is the parameter matrix in the GCN and $ \mathcal{D}_{s_{i}}$ is the neighbour distribution.

Moreover, for any positive scalar $t$, the likelihood that the Euclidean distance between the actual output $\mathbf{h}_i$ and this expected output exceeds $t$ is upper-bounded by:

\begin{equation}
     \mathbb{P} \left( \left\| \mathbf{h_i} - \mathbb{E} \left[ \mathbf{h_i} \right] \right\|_{2} \geq t \right) \leq 2 \cdot l \cdot \exp \left( -\frac{\operatorname{deg}(v_i) t^{2}}{2 \rho^{2}(\mathbf{W}) B^{2} l} \right)
\end{equation}
where $ l$ denotes the feature dimensionality and $\rho(\mathbf{W})$ denotes the largest singular value of $\mathbf{W}$.
\end{theorem}

\begin{proof}

A single GCN operation is defined by $\mathbf{H'} = \mathbf{D^{-1}}\mathbf{A}\mathbf{H}\mathbf{W}$, where $\mathbf{H}$ represents the input features and $\mathbf{H'}$ represents the output features of a given layer. $\mathbf{W}$ is a parameter matrix of size $l \times l$ that is responsible for the transformation of the features. Additionally, $D$ is a diagonal matrix, with its diagonal elements $D[i,i]$ equal to $\deg(i)$, which represents the degree of node $v_i$.

Focusing on a specific node $v_i$, the expectation of $\mathbf{h_i}$ can be derived as follows:

\begin{align}
\mathbb{E}\left[\mathbf{h_i}\right] &= \mathbb{E}\left[\sum_{j\in\mathcal{N}(v_i)}\frac{1}{\operatorname{deg}(v_i)} \mathbf{W} x_j\right] \\
&= \frac{1}{\operatorname{deg}(v_i)}\sum_{j\in\mathcal{N}(v_i)} \mathbf{W} \mathbb{E}_{c\sim\mathcal{D}_{s_i}, x\sim\mathcal{F}_b}[x] \\
&= \mathbf{W}\left(\mathbb{E}_{c\sim\mathcal{D}_{s_i}, x\sim\mathcal{F}_b}[x]\right).
\end{align}

Let $( \mathbf{x}_i[k], k=1,\ldots, l )$ denote the $i$-th element of  $x$. Then, for any dimension $k$, $\left\{ x_{j}[k], j\in\mathcal{N}(v_i) \right\}$ is a set of independent bounded random variables. Hence, directly applying Hoeffding's inequality, for any $t_{1} \geq 0$, we have the following bound:

\begin{equation}
    \mathbb{P}\left(\left|\frac{1}{\mathcal{N}(v_i)}\sum_{j\in \mathcal{N}(v_i)}\left(\mathbf{x}_{j}[k]-\mathbb{E}\left[\mathbf{x}_{j}[k]\right]\right)\right|\geq t_{1}\right) \leq 2\exp\left(-\frac{(\operatorname{deg}(v_i)) t_{1}^{2}}{2 \cdot B^{2}}\right)
\end{equation}

If  $\left\|\frac{1}{\mathcal{N}(v_i)} \sum_{j \in \mathcal{N}(v_i)}\left(\mathbf{x}_{j}-\mathbb{E}\left[\mathbf{x}_{j}\right]\right)\right\|_{2} \geq \sqrt{l} t_{1}$ , then at least for one  $k \in\{1, \ldots, l\}$ , the inequality  $\left|\frac{1}{\mathcal{N}(v_i)} \sum_{j \in \mathcal{N}(v_i)}\left(\mathbf{x}_{j}[k]-\mathbb{E}\left[\mathbf{x}_{j}[k]\right]\right)\right| \geq t_{1}$ holds. Hence, we have

\begin{align}
&\mathbb{P}\left(\left\|\frac{1}{\mathcal{N}(v_i)}\sum_{j\in\mathcal{N}(v_i)}\left(x_{j}-\mathbb{E}\left[x_{j}\right]\right)\right\|_{2}\geq\sqrt{l} t_{1}\right) \\
&\leq P\left(\bigcup_{k=1}^{l}\left\{\left|\frac{1}{\mathcal{N}(v_i)}\sum_{j\in\mathcal{N}(v_i)}\left(x_{j}[k]-\mathbb{E}\left[x_{j}[k]\right]\right)\right|\geq t_{1}\right\}\right) \\
&\leq\sum_{k=1}^{l} P\left(\left|\frac{1}{\mathcal{N}(v_i)}\sum_{j\in\mathcal{N}(v_i)}\left(x_{j}[k]-\mathbb{E}\left[x_{j}[k]\right]\right)\right|\geq t_{1}\right) \\
&=2\cdot l\cdot\exp\left(-\frac{(\operatorname{deg}(v_i)) t_{1}^{2}}{2  \cdot B^{2}}\right).
\end{align}

Let  $t_{1}=\frac{t_{2}}{\sqrt{l}}$ , then we have

\begin{equation}
    \mathbb{P}\left(\left\|\frac{1}{\mathcal{N}(v_i)} \sum_{j \in \mathcal{N}(v_i)}\left(\mathbf{x}_{j}-\mathbb{E}\left[\mathbf{x}_{j}\right]\right)\right\|_{2} \geq t_{2}\right) \leq 2 \cdot l \cdot \exp \left(-\frac{(\operatorname{deg}(v_i)) t_{2}^{2}}{2 \cdot B^{2} l}\right).
\end{equation}

Furthermore, we have

\begin{align}
\left\|\mathbf{h}_{i}-\mathbb{E}\left[\mathbf{h}_{i}\right]\right\|_{2} & =\left\|\mathbf{W}\left(\frac{1}{\mathcal{N}(v_i)} \sum_{j \in \mathcal{N}(v_i)}\left(\mathbf{x}_{j}-\mathbb{E}\left[\mathbf{x}_{j}\right]\right)\right)\right\|_{2} \\
& \leq\|\mathbf{W}\|_{2}\left\|\frac{1}{\mathcal{N}(v_i)} \sum_{j \in \mathcal{N}(v_i)}\left(\mathbf{x}_{j}-\mathbb{E}\left[\mathbf{x}_{j}\right]\right)\right\|_{2} \\
& =\rho(\mathbf{W})\left\|\frac{1}{\mathcal{N}(v_i)} \sum_{j \in \mathcal{N}(v_i)}\left(\mathbf{x}_{j}-\mathbb{E}\left[\mathbf{x}_{j}\right]\right)\right\|_{2}.
\end{align}
where  $\|\mathbf{W}\|_{2}$ refers to the L2 norm of matrix $\mathbf{W}$, which is the largest singular value of matrix $\mathbf{W}$. Additionally, the expression utilizes the identity that the L2 norm of matrix $\mathbf{W}$ is equal to its spectral radius $\rho(\mathbf{W})$. The spectral radius is the maximum absolute value of all the eigenvalues of matrix $\mathbf{W}$. 

Then, for any  $t>0$, we have

\begin{align}
&\mathbb{P}\left(\left\|\mathbf{h}_{i}-\mathbb{E}\left[\mathbf{h}_{i}\right]\right\|_{2} \geq t\right) \\& \leq \mathbb{P}\left(\rho(\mathbf{W})\left\|\frac{1}{\mathcal{N}(v_i)} \sum_{j \in \mathcal{N}(v_i)}\left(\mathbf{x}_{j}-\mathbb{E}\left[\mathbf{x}_{j}\right]\right)\right\|_{2} \geq t\right) \\
& =\mathbb{P}\left(\left\|\frac{1}{\mathcal{N}(v_i)} \sum_{j \in \mathcal{N}(v_i)}\left(\mathrm{x}_{j}-\mathbb{E}\left[\mathbf{x}_{j}\right]\right)\right\|_{2} \geq \frac{t}{\rho(\mathbf{W})}\right) \\
& \leq 2 \cdot l \cdot \exp \left(-\frac{(\operatorname{deg}(v_i)) t^{2}}{2 \rho^{2}(\mathbf{W}) B^{2} l}\right).
\end{align}

which completes the proof.

\end{proof}

\subsection{Independence}
\label{proof2}

\begin{theorem}
Let $p_{S}$ and $p_{\hat{Y}}$ represents the marginal density function of the random variable $S$ and $\hat{Y}$. $p_{\hat{Y}|S}$ is the conditional density function of $\hat{Y}$ given $S$, and $p_{\hat{Y}, S}$ is the joint density function of the random variables $\hat{Y}$ and $S$. Now, we introduce a discriminator $D$ which will discriminate between the model outputs $\hat{Y} = C(\mathbf{H})$ and whether the sensitive attribute $S$ is independent or not. The optimal discriminator $D^{*}$ that maximizes a certain objective function $\mathcal{L}_{in}$ over all possible discriminators $D$ can be expressed as:

\begin{equation}
p_{\hat{Y}, S}(\hat{y}, s)=p_{\hat{Y}}(\hat{y}) p_{S}(s).
\end{equation}
where $\hat{y} \in \hat{Y}$ and $s \in S$.

\end{theorem}

\begin{proof}

Let  $\hat{y} = C(\mathbf{h})$ where $\mathbf{h}$ is a specific node representation and $s$ is its sensitive attribute. The loss function $\mathcal{L}_{in}$ can be written:

\begin{align}
 &\int\log D(\hat{y}, s) p(\mathbf{h}, s) \, d\hat{y} \, ds + \int\log\left(1-D\left(\hat{y}, s^{\prime}\right)\right) p\left(\hat{y}, s^{\prime}\right) \, d\hat{y} \, ds^{\prime},\\
&=\int\log D(\hat{y}, s) p(\hat{y}\mid s) p(s)+\log(1-D(\hat{y}, s)) p(\hat{y}) p(s) \, d\hat{y} \, ds.
\end{align}

By the proof of Proposition 1 in ~\cite{goodfellow2014generative}, $\mathcal{L}_{in}$ is maximized at:
\begin{equation}
    D^*(\hat{y}, s)=\frac{p(\hat{y} \mid s) p(s)}{p(\hat{y} \mid s) p(s)+p(\hat{y}) p(s)}=\frac{p(\hat{y} \mid s)}{p(\hat{y} \mid s)+p(\hat{y})},
\end{equation}
for any  $\hat{y} \in \hat{Y}$  and $ s \in S $.

According to the argument in Theorem 1 of ~\cite{goodfellow2014generative}, $\mathcal{L}_{in}$ can be explained by the Jensen-Shannon divergence $JSD(\cdot,\cdot)$, i.e:

\begin{equation}
\mathcal{L}_{in} \left(C; D^{*}\right) = 2 J\left(P(\hat{Y}, S), P(\hat{Y}) P(S)\right) - \log 4
\end{equation}

If $JSD = 0$, it implies $p_{\hat{Y}, S}(\hat{y}, s)=p_{\hat{Y}}(\hat{y}) p_{S}(s)$ for $\hat{Y}$ and $S$. 
\end{proof}

\subsection{Separation}
\label{proof3}

\begin{theorem}

Let $p_{\hat{Y}}$ be the conditional density function of $\mathbf{H}$ given $Y$ and $p_{C(\mathbf{H}\mid S,Y}$ be of given $Y$ and $S$. Now, we introduce a discriminator $D$ which discriminates whether the model output $\hat{Y} = \hat{Y}$ is independent of the sensitive attribute $S$, given $Y$.The optimal discriminator $D^{*}$ that maximizes a certain objective function $\mathcal{L}_{se}$ over all possible discriminators $D$ can be expressed as:
    
\begin{equation}
\frac{p_{\hat{Y} \mid S, Y}(\hat{y} \mid s, y)}{p_{\hat{Y} \mid S, Y}(\hat{y} \mid s, y)+\epsilon(s, y) p_{\hat{Y} \mid Y}(\hat{y} \mid y) \frac{p_{S^{\prime}, Y}(\hat{y}, y)}{p_{S, Y}(\hat{y}, y)}},  
\end{equation}
for all $\hat{y} \in \hat{Y}$, $s \in S$, and $y \in Y$, where $p_{S',Y}$ and $p_{S,Y}$ be the joint density functions of $S'$ and $Y$ and of $S$ and $Y$ respectively.
\end{theorem}

\begin{proof}
    Let  $\hat{y} = C(\mathbf{h})$ where $\mathbf{h}$ is a specific node representation and $s$ is its sensitive attribute. The loss function $\mathcal{R}_{se}$ can be written:
\begin{align}
&\mathcal{R}_{se} =E_{\mathbf{H}, S, Y}[\log D(\hat{Y}, S, Y)]+ \notag\\
& E_{S^{\prime}} E_{\mathbf{H}, Y}\left[\epsilon\left(S^{\prime}, Y\right)\log\left(1-D\left(C(\mathbf{H}, S^{\prime}, Y\right)\right)\right],\\
&=\int\log D(\hat{y}, s, y) p(\hat{y}\mid s, y) p(s, y) \, ds \, dy \, d\hat{y} + \notag \\
& \int\epsilon\left(s^{\prime}, y\right)\log\left(1-D\left(\hat{y}, s^{\prime}, y\right)\right) p(\hat{y}\mid y) p\left(s^{\prime}\right) p(y) \, ds^{\prime} \, dy \, d\hat{y},\\
&=\int\log D(\hat{y}, s, y) p(\hat{y}\mid s, y) p(s, y)+ \notag\\
&\epsilon(s, y)\log(1-D(\hat{y}, s, y)) p(\hat{y}\mid y) p(s) p(y) \, ds \, dy \, d\hat{y}.
\end{align}

By the proof of Proposition 1 in ~\cite{goodfellow2014generative}, $R_{se}$ is maximized at:
\begin{align}
&D^*(\hat{y}, s, y;\epsilon)\\
&=\frac{p(\hat{y}\mid s, y) p(s, y)}{p(\hat{y}\mid s, y) p(s, y)+\epsilon(s, y) p(\hat{y}\mid y) p(s) p(y)}\\
&=\frac{p(\hat{y}\mid s, y)}{p(\hat{y}\mid s, y)+\epsilon(s, y) p(\hat{y}\mid y)\frac{p(s) p(y)}{p(s, y)}}\\
&=\frac{p_{\hat{Y} \mid S, Y}(\hat{y} \mid s, y)}{p_{\hat{Y} \mid S, Y}(\hat{y} \mid s, y)+\epsilon(s, y) p_{\hat{Y} \mid Y}(\hat{y} \mid y) \frac{p_{S^{\prime}, Y}(\hat{y}, y)}{p_{S, Y}(\hat{y}, y)}}.    
\end{align}

\end{proof}

\end{document}